\documentclass[12pt]{article}
\newcommand{\f}{\begin{equation}}
\newcommand{\ff}{\end{equation}}
\newcommand{\fa}{\begin{eqnarray}}
\newcommand{\ffa}{\end{eqnarray}}
\usepackage[dvips]{graphicx}
\title{Introduction to supersymmetric spin networks}
\author{ Yi Ling \thanks{email:
$^{*}$ ling@phys.psu.edu}\\ \centerline{\it Center for
Gravitational Physics and Geometry}\\ \centerline{\it Department
of Physics}\\ \centerline {\it The Pennsylvania State
University}\\ \centerline{\it University Park, PA, USA 16802}\\
\centerline{and} \\ \centerline{\it The Blackett Laboratory}  \\
\centerline{\it Imperial College of Science, Technology and
Medicine}\\ \centerline{\it South Kensington, London SW7 2BZ, UK}}
\begin{document}
\maketitle
\begin{abstract}
\baselineskip=20pt In this paper we give a general introduction to
supersymmetric spin networks. Its construction has a direct
interpretation in context of the representation theory of the
superalgebra. In particular we analyze a special kind of spin
networks with superalgebra $Osp(1|2n)$. It turns out that the set
of corresponding spin network states forms an orthogonal basis of
the Hilbert space $\cal L\mit^2(\cal A\mit/\cal G)$, and this
argument holds even in the q-deformed case. The $Osp(n|2)$ spin
networks are also discussed briefly. We expect they could provide
useful techniques to quantum supergravity and gauge field theories
from the point of non-perturbative view.
\end{abstract}
\section{Introduction}
\baselineskip=20pt

The notion of spin networks originally was advocated by Roger
Penrose in 1970s when he tried to give a quantum mechanical
description of the geometry of space\cite{penrose}. In his
opinion, the final version of quantum geometry should be a
combinatorial theory in which we consider the different
combinations and permutations of objects such that we could derive
the discrete spectra of observables in the quantum mechanical
level. After that the idea of spin networks was introduced to many
areas, including lattice gauge theory\cite{sn-gauge} and
topological field theory\cite{sn-topo}. In the middle of 1990s,
the spin networks was introduced in loop quantum gravity in a
quite different way\cite{sn-qg, rigorous}. It was exploited to
construct the Hilbert space of kinematical quantum states and
consequently the discrete spectra of the area and volume of the
space were obtained\cite{sn1}. Later the dynamics of the spin
networks were also considered and its evolution gives rise to
casual set of spin networks\cite{fotini} or spin foams\cite{foam}.

It's evident to see the importance of spin networks if we list
some basic features it contains. First spin networks is a very
general notion in quantum field theory in which gauge fields
involve. In particular they are gauge invariant objects, in the
sense that the corresponding spin network states will be solutions
to the Gauss constraint naturally if we take the standard Dirac
procedures to quantize the theory. As a result, it would be much
easier to find the physically related subspace in Hilbert space.
In path integral formulation, we can consider the functional
integration on the modular space since a well-defined measure
theory can be established in context of spin networks. In loop
quantum gravity, they are background independent and
non-perturbative objects as well.

Until now we mainly focus on $SU(2)$ spin networks since it has
important application to quantum general relativity\cite{sn1,foam}
\footnote{ We also notice that spin networks with other Lie groups
have also been discussed in some places\cite{othersn}.}. However,
in principle we could construct the spin networks associated to
other groups or supergroups. More importantly we have a belief
that they could be applied to quantum supergravity and gauge
theories as well. Based on this motivation, the supersymmetric
spin networks firstly was introduced in \cite{superspin} and then
was developed in \cite{supern=2}. In \cite{superspin}by virtue of
supersymmetric spin networks, we carry out a non-perturbative
quantization of simple supergravity, in particular we find the
spectrum of area operator taking a discrete form, \f
\hat{A}|\Gamma^{sg},e_i,v_j\rangle=\sum_i l_{p}^2
\sqrt{j_i(j_i+\frac{1}{2})}| \Gamma^{sg},e_i,v_j\rangle,
\label{area}\ff where $l_{p}$ is the Planck length and
$j_i={n_i\over 2}$.

In this paper we develop the construction of supersymmetric spin
networks. After giving an overview on some general features of
spin networks in section two, we study a special kind of super
spin networks which has superalgebras $Osp(1|2n)$ in the
consequent section, and $Osp(N|2)$ spin networks is also discussed
briefly in section four. In section five we discuss some possible
applications of super spin networks.
\section{An overview }

Spin network is a graph, $\Gamma(e_i, v_j)$, embedded in a three
dimensional manifold $M$, which is composed of edges (or links)
and vertices (either nodes or joints). To each edge, we assign the
color, $e_i$, which is related to the labels of the irreducible
representation of groups, and to each vertex, we assign an
intertwiner operator, $v_j$, which maps the incoming irreducible
representations to the outcoming ones at the vertex.\footnote{we
adopt the convention that the vertex will be called k-valent one
if there are $k$ edges meeting at this vertex.}

The construction of spin networks has a direct interpretation in
representation theory of groups. In spin networks, each edge $e$
labeled by the representation could be understood as a parallel
propagator or holonomy of the connection $A$, $U_e(A)$, along the
edge in connection representation. In matrix notation of group
theory, it also corresponds to the higher dimensional irreducible
representation of the group element. In $SU(2)$ spin networks, we
also think of the edge as the combination of many {\em ropes},
each of which corresponds to the fundamental presentation of the
group. Therefore the reason that each edge can be decomposed into
many ropes stems from the fact that every higher dimensional
irreducible representation of the group can be obtained by
employing the symmetrization or anti-symmetrization procedures
from the fundamental representations. It is this fact that we can
decompose the spin network into multi-loop graphs by permuting and
connecting all the ropes to form loops, and finally we are able to
establish the transformation between spin network states and loop
states in the corresponding Hilbert space.

Intertwiner operators associated with every vertex in spin network
can be understood as the different ways that we could carry out to
connect the ropes when edges meet at the same vertex.
Correspondingly, in the language of representation theory of
groups, it corresponds to the fact that tensor products of several
irreducible representations can be completely decomposed into the
direct sum of the irreducible representations. In hence they are
invariant tensors in irreducible representations of groups and
given by standard Clebsch-Gordan $(CG)$ theory. In the case of
$SU(2)$ spin networks, when the vertex is a tri-valent one, the
decomposition of the tensor product is unique. If the edges are
more than three, we then can divide the multi-valent vertex into
the tri-valent vertices by making use of the intertwiner operator.
At the same time, restricted by the expansion of Clebsch-Gordan
series, the colors associated with edges which meet at the same
vertex must satisfy some conditions consistent with these $CG$
series. We call them admissible conditions. For instance, consider
three valences with colors $(a, b, c)$ meet at the same vertex in
ordinary $SU(2)$ spin networks, then they have to satisfy the
triangle inequality and the sum of them has to be even numbers,
however, in the case of $Osp(1|2)$ then the sum can be any
positive integer.

Associated with each spin network, we can obtain one number by
taking the trace of corresponding matrix product of the
propagators along edges in representation space, which is called
the evaluation of spin networks. It has very important
applications to quantum gravity. In particular when we consider
the action of operators such as area and volume observables on the
spin network states, it provides us a practical way to work out
the spectra of these observables. In cases of $SU(2)$ and
$Osp(1|2)$, the evaluations of spin networks, in particular the
theta graphs, are discussed respectively in \cite{KL,superspin},
where $6j$ symbols and recoupling theories play important roles.

Based on the spin network $\Gamma(e_i, v_j)$, we can define a spin
network state, $\Phi_{f_\Gamma}(A)$, by means of the cylindrical
function with the form \f
\Phi_{f_\Gamma}(A):=f_{\Gamma}(U_{e_1}(A),...,U_{e_n}(A))\label{sns},\ff
where cylindrical function $f_{\Gamma}$ refers to taking the
holonomy along each edge and then contracting the holonomy matrix
with the intertwiners at each vertex where edges meet. Spin
network states have more advantages than loop states since they
are linear independent and do form a basis of the Hilbert space,
rather than loop states the space of which is over completed and
contains some identities. To show the spin network states form a
basis in the Hilbert space, We need solve two key problems. One is
the definition of inner product of the spin network state,
consequently we can show any two different spin network states are
orthogonal and linear independent; the other one is the
completeness of the spin network states, namely any state in
Hilbert space can be expressed in terms of spin network states. In
the case of $SU(2)$, these two problems are solved successfully
mainly by virtue of the Haar measure and Peter-Weyl theorem in
group theory respectively \cite{sn-qg, rigorous, sn1}. As a
result, $SU(2)$ spin networks plays a key role to form a linear
independent basis of the Hilbert space in loop quantum gravity.

When we try to extend the notion of spin networks to the
supersymmetric case, we need construct the graphs and find the
rules which must be completely consistent with the representation
theory of superalgebras. Following constructions of $SU(2)$ and
$Osp(1|2)$ spin networks\cite{sn-qg,superspin} which has been
recalled above, let us list some basic procedures that we have to
take into account in this paper.
\begin{itemize}
\item
The definition of supersymmetric spin networks. At the first
sight, it's simple to define the supersymmetric spin networks. we
only need to change the representations associated to edges to the
corresponding representations of the supergroups and label the
vertex by intertwiner operator appropriately. But after that, to
make the spin network well defined, we need to consider the
following related questions.
\item
What's the admissible conditions associated with the trivalent
vertex? Or equivalently, can any tensor product of the irreducible
representations be decomposed into the direct sum of the
irreducible representations? Due to the features of superalgebras
their own, we will face some troubles at once, because unlike the
Lie algebra, {\em the tensor products of many kinds superalgebras
are not completely reducible into the irreducible ones.} We could
see this trouble elsewhere when we construct the $Osp(2|2)$ spin
networks. \cite {supern=2}
\item
Could we find a way to evaluate the graphs such that we could
consider the action of the operators on the corresponding spin
network states and then calculate the spectra of the operators?
More explicitly, can every edge be decomposed into ropes, as we do
in the case of $SU(2)$? Namely, can any irreducible representation
of the supergroups be constructed from the fundamental
representation? We will also see only some special sorts of
superalgebras have such features. Furthermore, is it possible to
carry out a graphic representation of computing the enclosure of
edges?
\item
Does the set of the corresponding spin network states form a basis
for the Hilbert space? To show this, first we need show that any
different spin network states are orthogonal and linear
independent, second any states in the space can be decomposed into
the sum of the spin network states.
\end{itemize}

To answer these questions, in this paper we extend the strategy in
$SU(2)$ spin networks to a special kind of supersymmetric spin
network which is equipped with superalgebra $Osp(1|2n)$.  The
analysis of this special one will lead to some general comments on
the construction of supersymmetric spin networks with other
superalgbras.

But before we do that, let us recall some basic facts related to
superalgebras in the last part of this section. Superalgebras and
supergroups were proposed in physics to construct the
supersymmetric model in which bosons and fermions are placed in
the same supermultiplet and they could change into each other
under the supersymmetric transformation. In fact, every
superalgebra or graded algebra contains two kinds of generators.
One is even which is the bosonic part and the other is odd
corresponding to fermionic part. Contrasting to the ordinary
algebras, the even generators of the superalgebra are associated
with the commuting parameters while the odd ones are associated
with the $anti$-commuting parameters. A special kind of
superalgebra is classical simple Lie superalgebra, whose odd part
is completely reducible into one or two irreducible subspace. Its
classification and representation is given in the fundamental
paper by Kac\cite{Kac}. Furthermore, if classical superalgebras
admit a nondegenerate metric tensor, it's called basic
superalgebras, which is the closest one to simple Lie algebra. All
irreducible representations of basic superalgebras are obtainable
from a highest weight and the Schur lemma holds under the usual
way. However, unlike the ordinary Lie algebras, normally
superalgebras have two kinds of representations. One is typical,
and the other one is atypical. The typical representations are
irreducible and more like the ordinary representations of the Lie
algebra, however the atypical ones are in many respects
degenerate. In particular the atypical representations maybe are
not {\it completely} reducible, for example if they occur as the
semidirect sums of several irreducible atypical representations.

Many kinds of Lie superalgebras do not have properties similar to
those of Lie algebras. For example, the complete reducibility is
not valid to all the simple Lie superalgebra any more, even to
basic ones. Finite reducible but indecomposable representations
may appear if we consider the tensor products of irreducible
representations. However, the representations of $Osp(1|2n)$ have
all the nice properties of those of semisimple Lie algebras. For
example, its all reducible representations are fully reducible and
therefore a generalized Wigner-Eckhart theorem holds.

The Lie supergroups are obtained by exponentiating Lie
superalgebras. Particularly for supergroups $SU(N|M)$ and
$Osp(N|2M)$, it's known that all the representations constructible
can be obtained from the direct product of fundamental
representations.

\section{$Osp(1|2n)$ spin networks}

\subsection{Definition}

\begin{figure}[h]
\begin{center}
\includegraphics[angle=0,width=6.5cm,height=2cm]{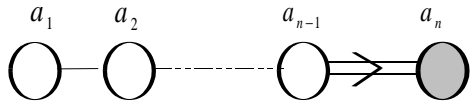}
\caption{ Dynkin diagram of $Osp(1|2n)$} \label{Dynkin}
\end{center}
\end{figure}

First let us concentrate on a special kind of spin networks with
superalgebra $Osp(1|2n)=B(0,n)$\cite{dic}, which is a subset of
the orthosympletic Lie superalgebras $Osp(M|2n)$ \footnote{The
classification of superalgebra was established in \cite{Kac}.
There are three kinds of orthosympletic algebras $Osp(M|2n)$:
$B(m,n)=Osp(2m+1|2n)$; $C(n)=Osp(2|2n-2),(n\geq 2)$ and
$D(m,n)=Osp(2m|2n), (m,n\geq 2)$.}. Its even part is
$O(1)\bigotimes Sp(2n)$ and the Dynkin diagram is shown in
Fig.(\ref{Dynkin}).
The finite dimensional irreducible representation is characterized
by its highest weight $\Lambda=(a_1,a_2,...a_n)$ which takes the
form \f \Lambda=a_1\omega_1+a_2\omega_2+...a_n\omega_n,\ff where
$\omega_i(i=1...n)$ is the fundamental weight and the coordinate
$a_i(i=1...n)$ has to be non-negative integer. The $Sp(2n)$
representation containing in the representation can be read out
from the diagram directly by replacing the odd root $a_n$ by
$b_n={1\over 2}a_n$. Correspondingly, we assign the edge in
$Osp(1|2n)$ spin networks is labeled by $e_i$ which is defined as
a partition of $(a_1, ..., a_n)$. In the language of spin networks
we also call the weight coordinates colors of the edge. In the
simplest case of $n=1$, we see the representation of superalgebra
is labeled by only one integer $e_1=2j_1$, and the corresponding
spin networks is discussed in detail in \cite{superspin}.
\subsection{Elements: edges and vertices}
\begin{figure}[h]
\begin{center}
\includegraphics[angle=0,width=8cm,height=4cm]{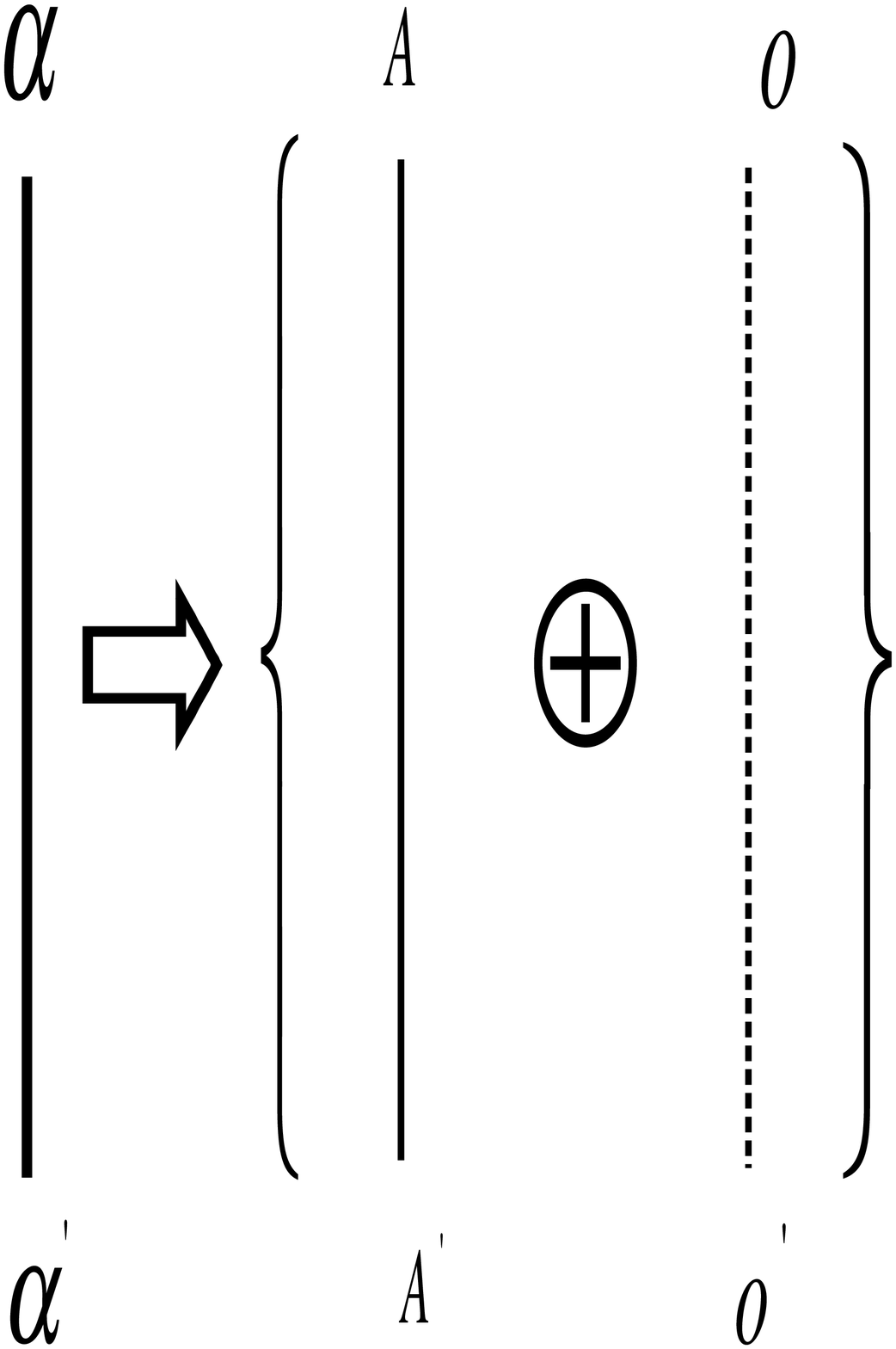}
\caption{Composition of unit element with color $(1,0...,0)$.}
\label{comp}
\end{center}
\end{figure}

As we mentioned above, in general there are two kinds of
representations of the superalgebra. One is typical, and the other
one is atypical. However, superalgebra $Osp(1|2n)$ has a
remarkable feature. It has only typical representation. The
fundamental representation is labeled by weight coordinates $(1,
0, ..., 0)$, and the basis vector of the representation is $2n+1$
dimensional and contains one boson and $2n$ fermions.\f
\xi_\alpha=\left(\begin{array}{c}
  \psi_A \\
  \phi_o
\end{array}\right),\ff
where $\psi_A$ is the fermionic part of the representation and
$A=(1,..., 2n)$ is the spinor index of $Sp(2n)$, while $\phi_o$ is
the bosonic part which is only one dimensional.  Then the unit
element in the fundamental representation can be illustrated as
fig.(\ref{comp}), in which we denote the $Sp(2n)$ element by the
thin line and the single bosonic part by the dotted line.

The higher finite dimensional representations of the superalgebra
can be obtained by symmetrizing and anti-symmetrizing the
fundamental representations\cite{GSU5}. Note that in the simplest
case of $n=1$, we only need take the symmetrization procedures
since the anti-symmetrization of two spinor indices will be
identical to the trivial representation. In context of Young
tableaus, we have only one row of boxes. However, for $n\geq 2$,
we need take both symmetrization and anti-symmetrization
procedures to obtain all the finite dimensional irreducible
representations. For instance, we consider the tensor products of
two fundamental representations, we can symmetrize two basis
vectors which is defined as,\footnote{Strictly speaking, unlike
the superalgebra $SU(N|M)$, the symmetrized basis
$\xi_{(\alpha\beta)}$ need to shift a little bit by the condition
$G_{\alpha\beta}\xi_{(\alpha\beta)}=0$\cite{GSU5}, where
$G_{\alpha\beta}$ is the invariant bilinear form taking the form
$\left(\begin{array}{ccc}
  1 & 0 & 0 \\
  0 & 0 & I_n \\
  0 & -I_n & 0
\end{array}\right)$. However this shift will not affect our
construction of the unit element of the supergroups. We just
ignore this shift in this paper.}

\begin{figure}[h]
\begin{center}
\includegraphics[angle=0,width=8cm,height=4cm]{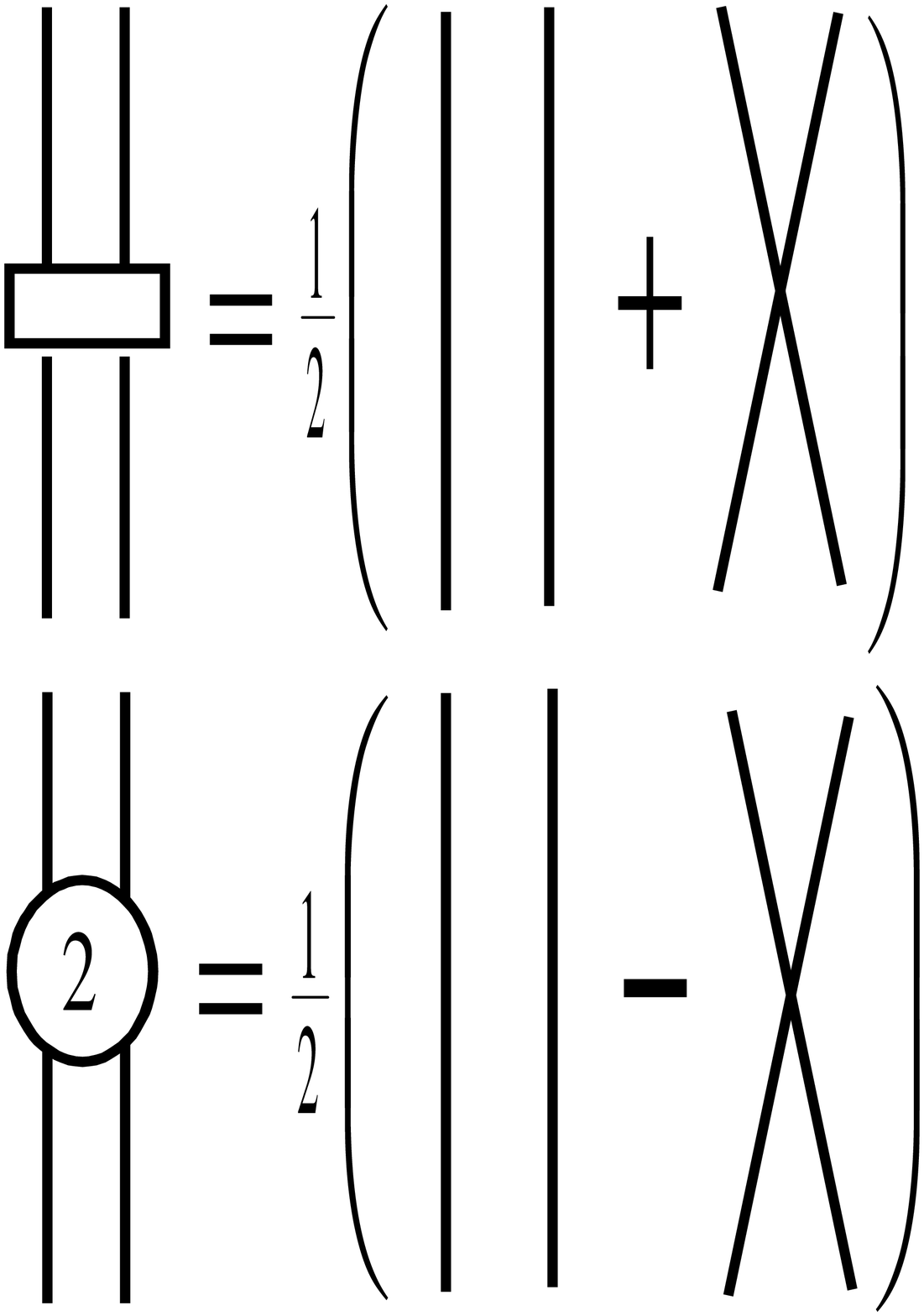}
\caption{Symmetrization and anti-symmetrization of fundamental
representation}\label{symm}
\end{center}
\end{figure}

\f \xi_{(\alpha\beta)}=\xi^1_\alpha\xi^2_\beta+(-)^{g(\alpha)\cdot
g(\beta)}\xi^1_\beta\xi^2_\alpha.\ff Also we can antisymmetrize
them as  \f
\xi_{(\alpha\beta)}=\xi^1_\alpha\xi^2_\beta-(-)^{g(\alpha)\cdot
g(\beta)}\xi^1_\beta\xi^2_\alpha,\ff where $g(\alpha)$ is the
grade of the index. For fermionic indices, it's one and for
bosonic it's zero. If we denote the symmetrization by square box
in the graph, and antisymmetrization by a circle labeled by the
number of ropes under consideration, then the unit elements of the
supergroup under these representations are illustrated in
Fig.(\ref{symm}).

\begin{figure}[h]
\begin{center}
\includegraphics[angle=0,width=6cm,height=5cm]{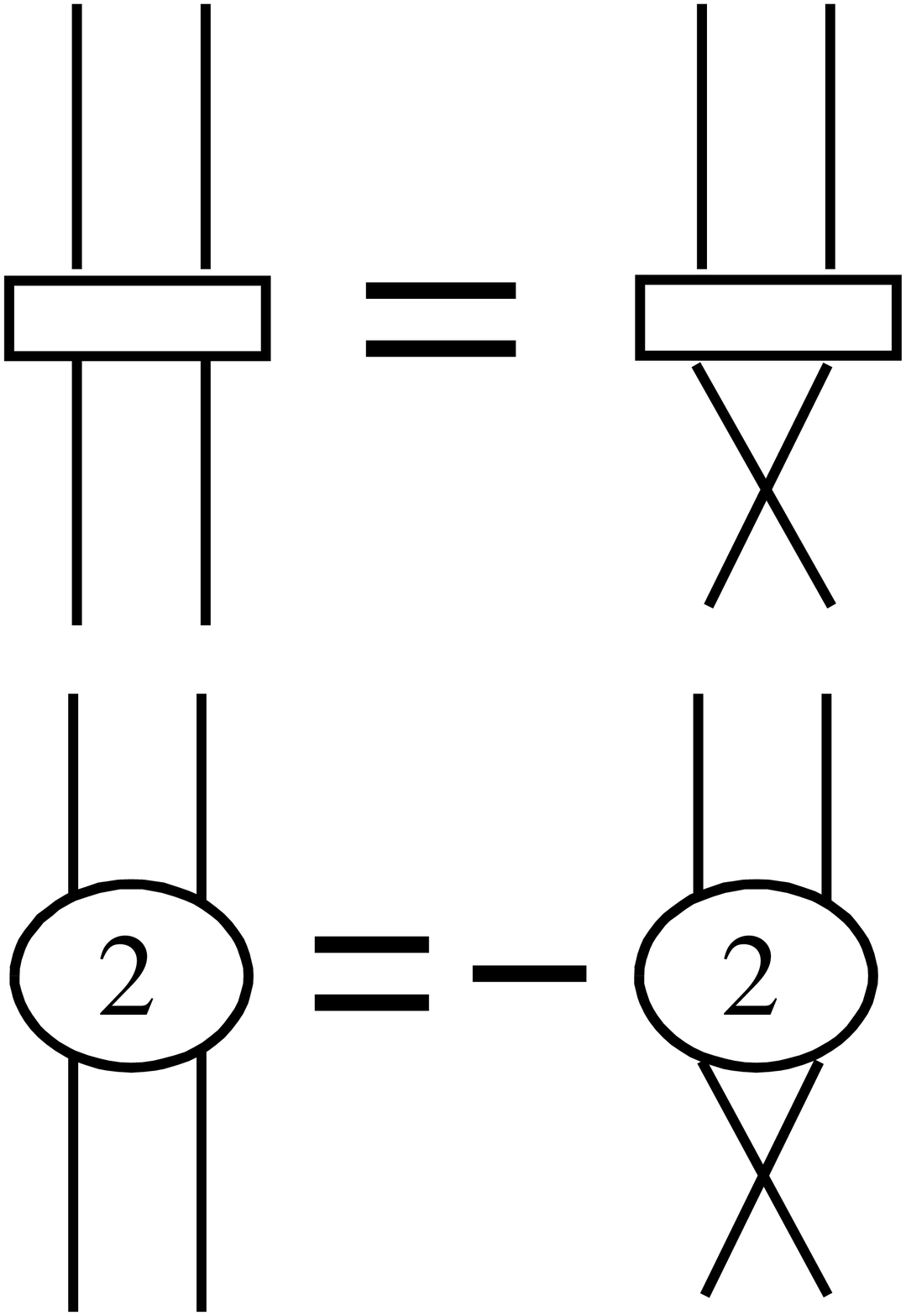}
\caption{Property I}\label{proI}
\end{center}
\end{figure}

\begin{figure}[h]
\begin{center}
\includegraphics[angle=0,width=6cm,height=4cm]{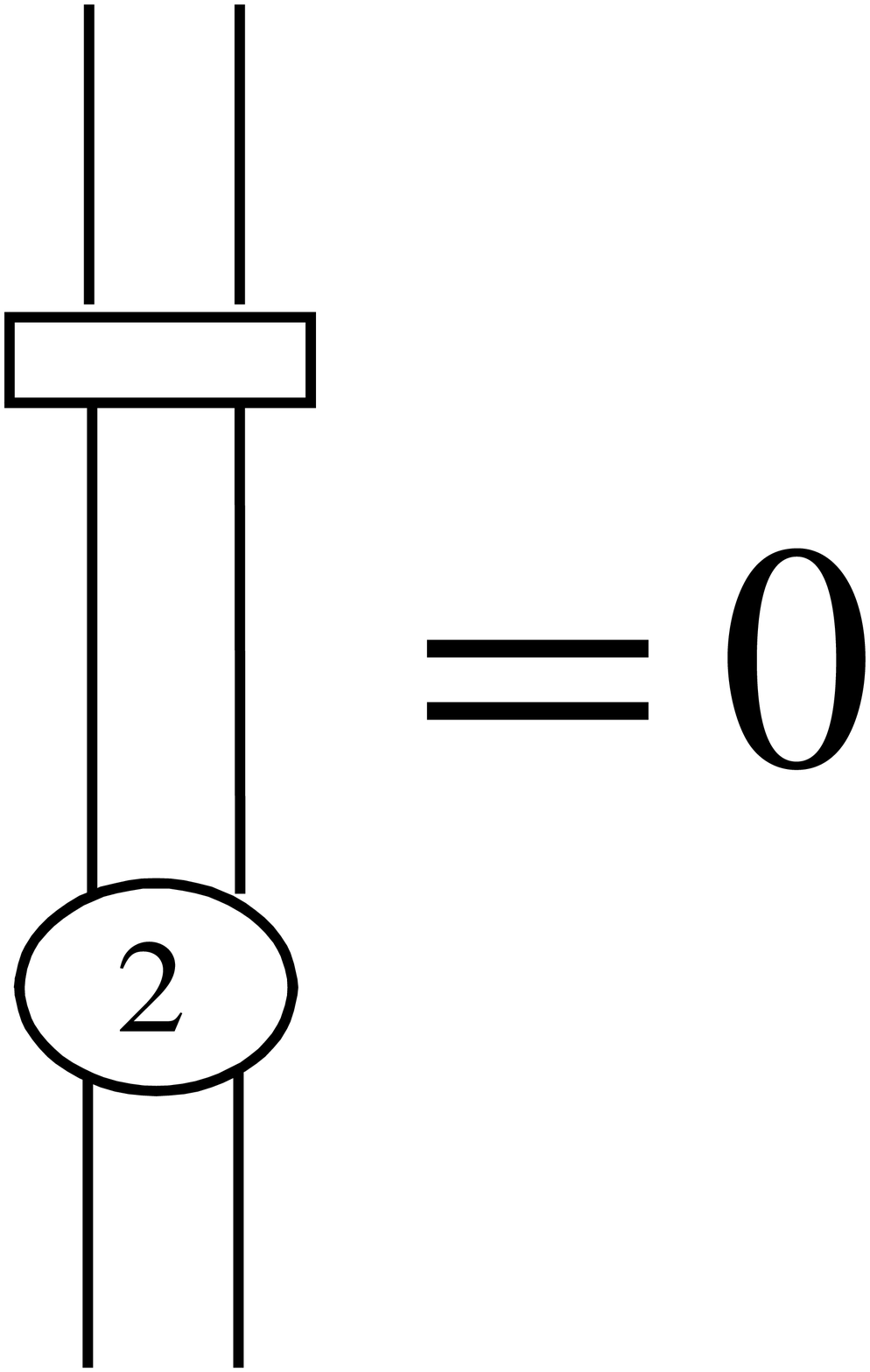}
\caption{Property II}\label{proII}
\end{center}
\end{figure}

\begin{figure}[h]
\begin{center}
\includegraphics[angle=0,width=8cm,height=4cm]{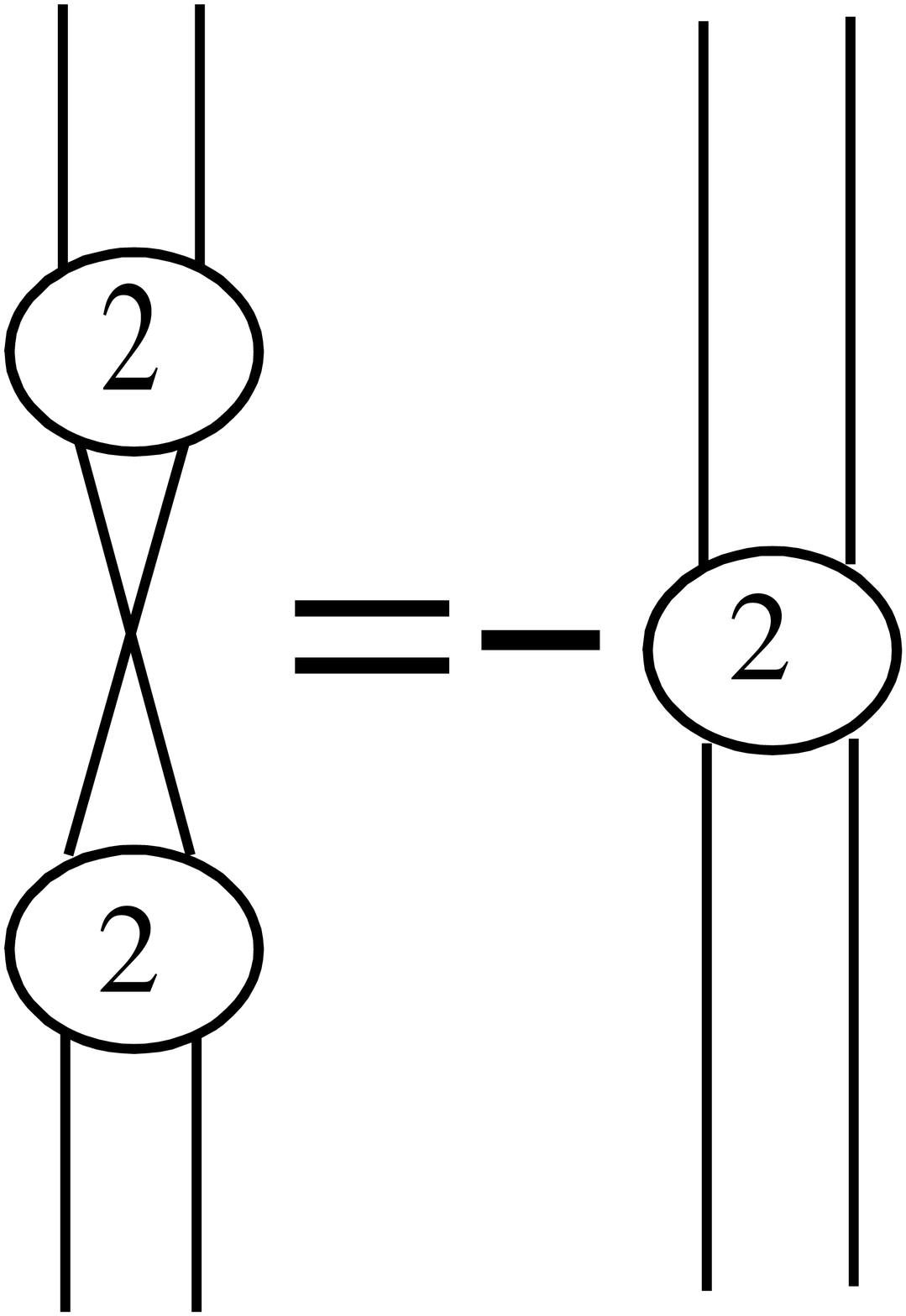}
\caption{Property III}\label{proIII}
\end{center}
\end{figure}

Their weight coordinates are $(2, 0, ...0)$ and $(0, 1, 0, ...,
0)$ respectively. We also call the unit element symmetrizer in
which ropes are symmetrized and antisymmetrizer in which ropes are
antisymmetrized. Unlike the symmetrization, We can derive the
following properties of the antisymmetrization in graphic
representations, see fig.(\ref{proI}), fig. (\ref{proII}) and fig.
(\ref{proIII}).

\begin{figure}[h]
\begin{center}
\includegraphics[angle=0,width=8cm,height=4cm]{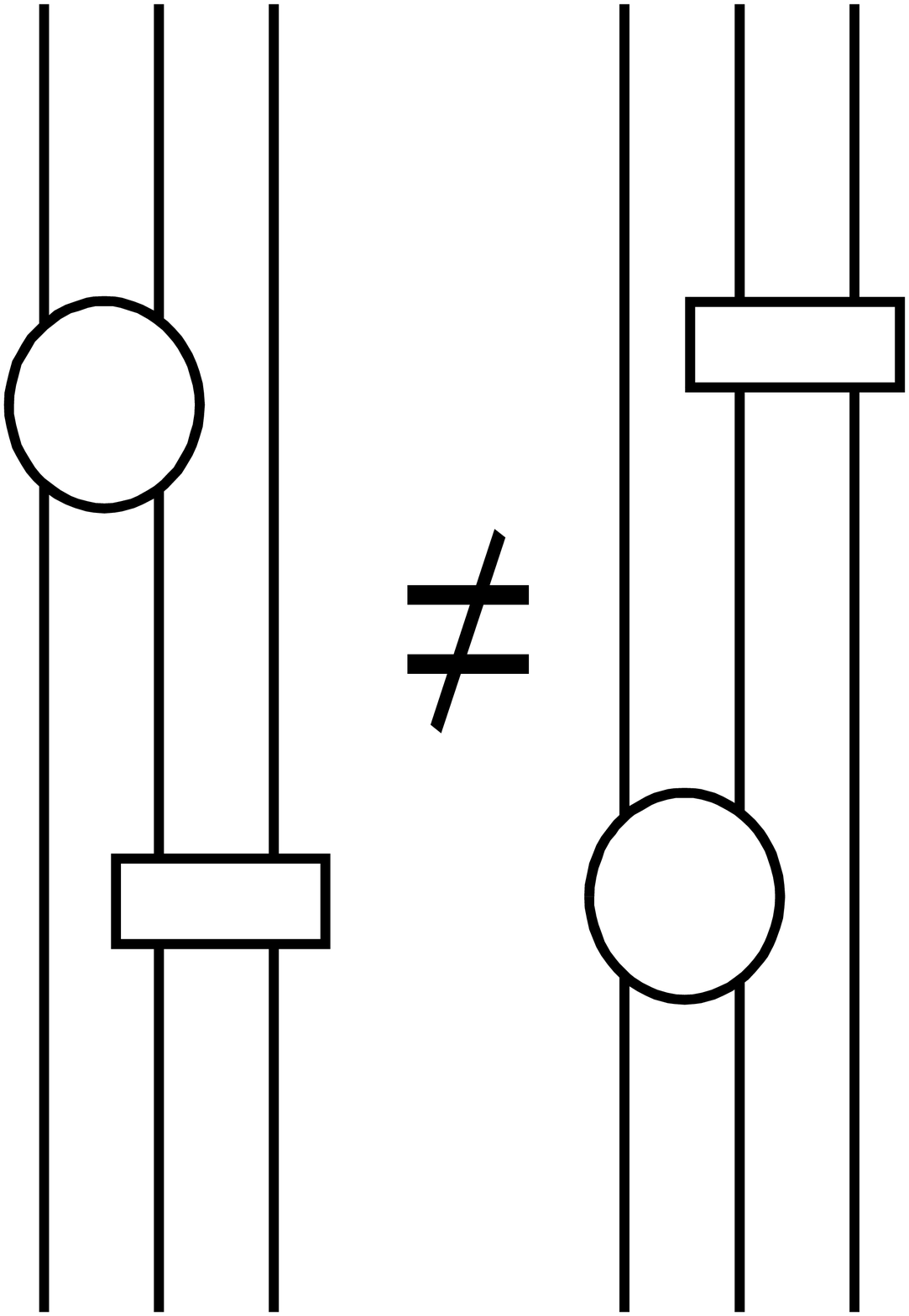}
\caption{Two distinguished graphs with different orders of box and
circle}\label{dis}
\end{center}
\end{figure}

These properties are easily proved by using the definition of
symmetrizers and antisymmetrizers shown in (\ref{symm}). For
instance in fig.(\ref{proII}), when two lines are both symmetrized
and antisymmetrized at the same time, then obviously they
vanishes.
Also when one graph involves both boxes and circles, we find the
order is also important now. For instance, the graphs shown in
fig.(\ref{dis}) are not equivalent.

\begin{figure}[h]
\begin{center}
\includegraphics[angle=0,width=8cm,height=4cm]{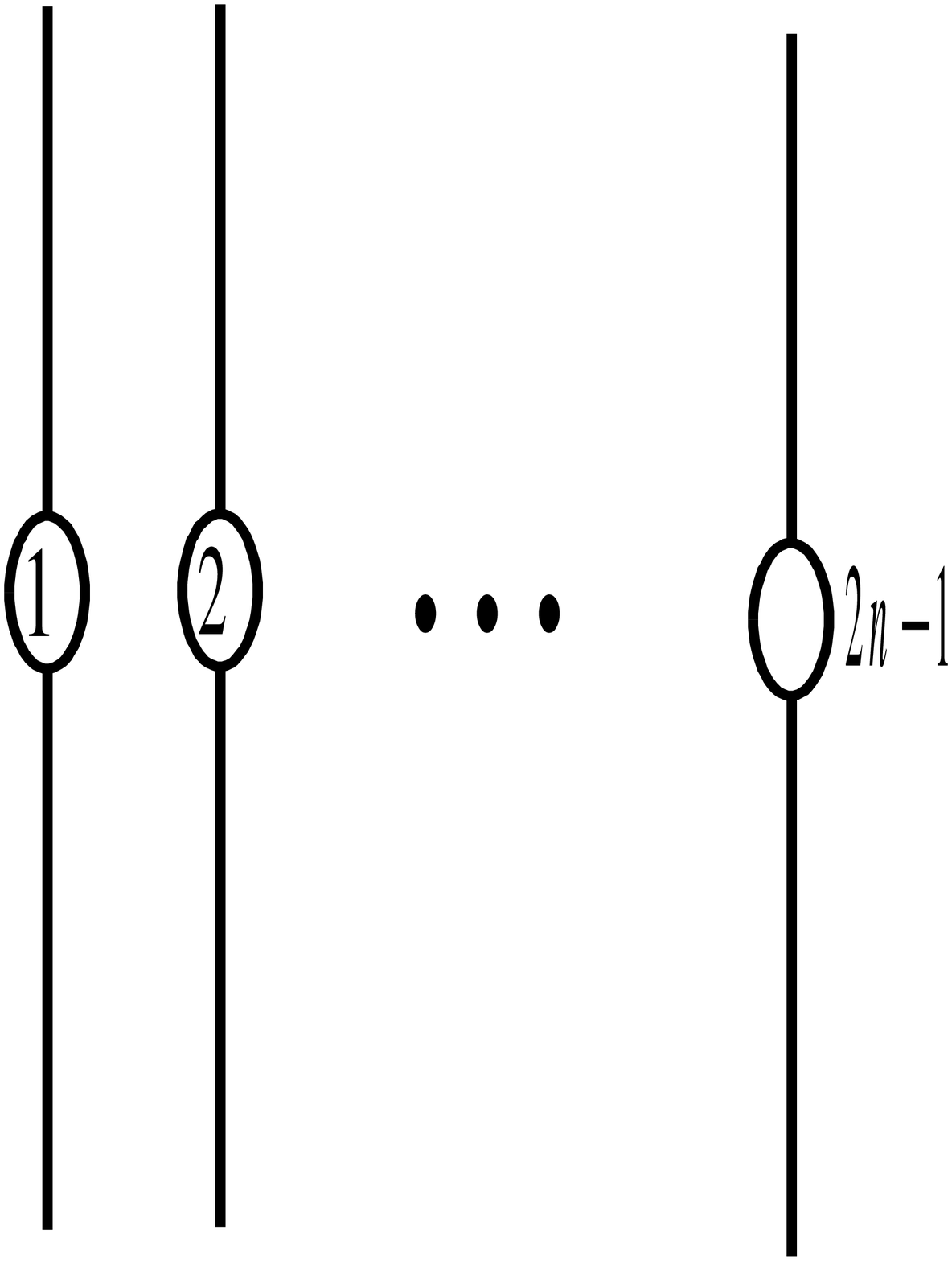}
\caption{All the non-vanishing antisymmetrizers in $Osp(1|2n)$
spin networks}\label{anti}
\end{center}
\end{figure}

Continuously taking the antisymmetrization procedures on
fundamental representations, we will find it has to be terminated
as far as more than $2n$ fundamental representations involve since
antisymmetrizing the same index vanishes. To simplify the
notation, we only show one line but label the number of ropes
which are anti-symmetrized in the circle. We draw it as
fig.(\ref{anti}).

To construct the graphical representation for any highest weight
irreducible representation, we need make use of the Young
supertableaux. For $Osp(1|2n)$, the construction of Young
supertableaux is simple since it has the same shape as the usual
Young tableaus of $Sp(2n)$ representation $(a_1, a_2, ...,
a_{n-1},{1\over 2}a_n)$. The Young tableau associated to $Sp(2n)$
representation $(a_1, a_2, ..., a_{n-1}, \frac{1}{2}a_n)$ is
defined as a graph with $\lambda_i$ boxes in the $i$th row where
$\lambda_i$ is related to the Dynkin labels by, \f
\lambda_i=a_i+\lambda_{j+1},\ \ \ \ \lambda_n={1\over 2}a_n.\ff
which satisfies $\lambda_1\geq\lambda_2\geq ...\geq\lambda_n\geq
0.$

\begin{figure}[h]
\begin{center}
\includegraphics[angle=0,width=8cm,height=4cm]{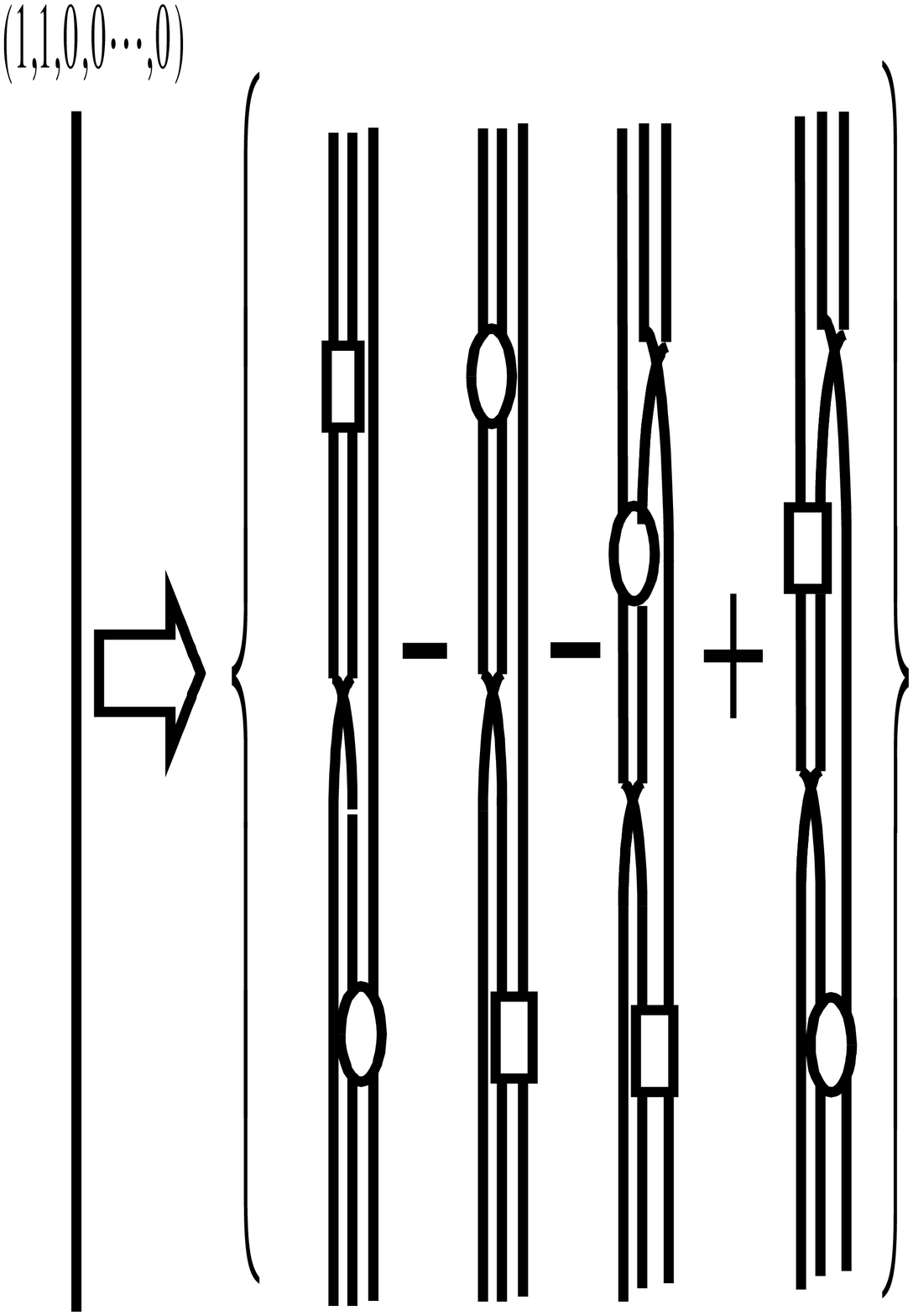}
\caption{The graphic representation of $(1, 1, 0,...,
0)$}\label{rep1}
\end{center}
\end{figure}

Replacing each box in Young supertableaux by a vertical straight
rope and putting them in parallel from left to right, then we
define a decomposition of the edge by symmetrizing or
antisymmetrize them corresponding to their positions in Young
supertableaux.  As a result, if an irreducible representation is
obtained by taking both symmetrization and anti-symmetrization
procedures, we find there are several inequivalent graphic
representations. For example, the representation $(1, 1, 0, 0,
..., 0)$ could be realized by graphs illustrated in
fig.(\ref{rep1}).

\begin{figure}[h]
\begin{center}
\includegraphics[angle=0,width=6cm,height=4cm]{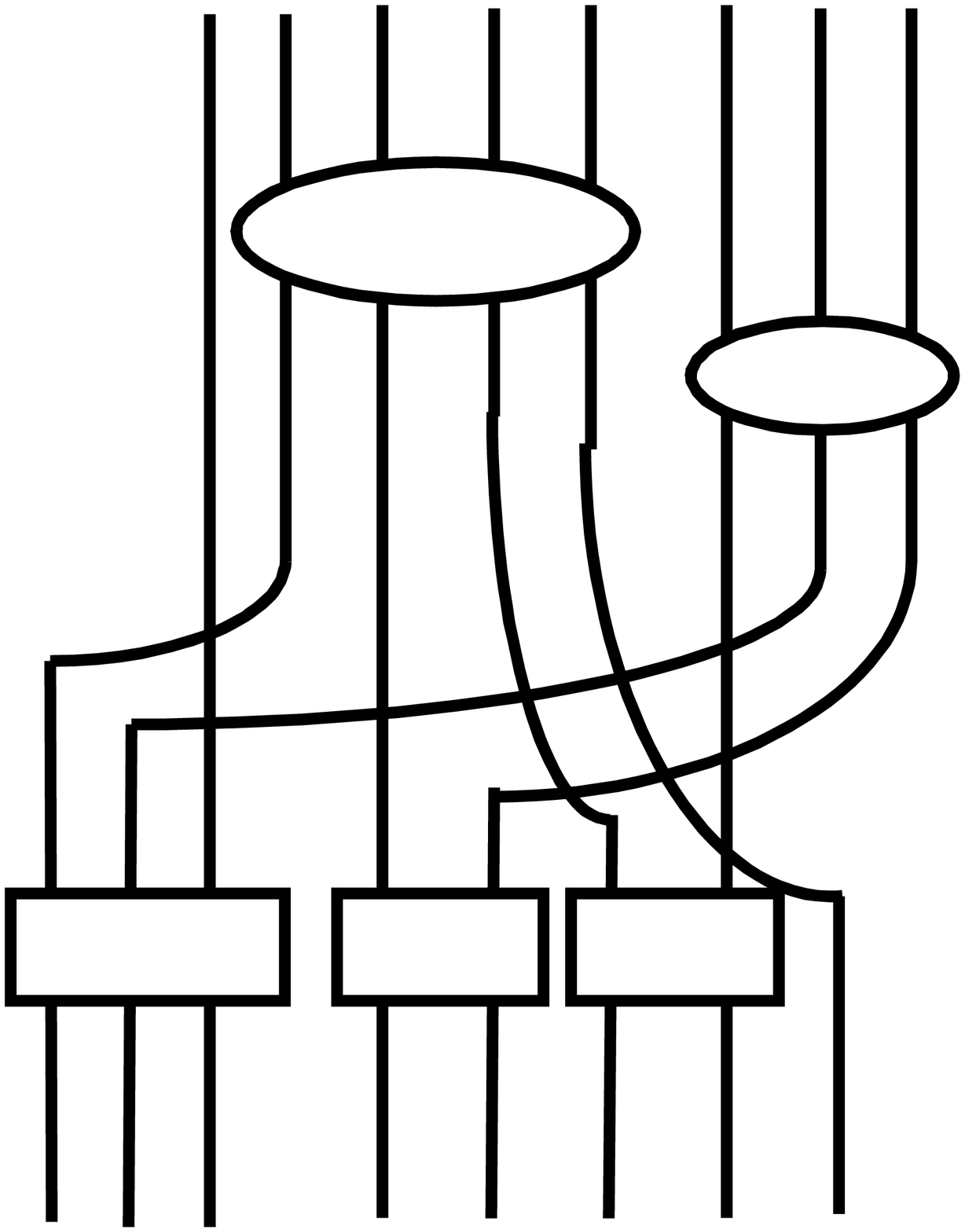}
\caption{The rope components of general edges} \label{edge}
\end{center}
\end{figure}

In conclusion we find the edge in spin networks can be decomposed
into some components which are the combination of some symmetrized
or antisymmetrized ropes. An example of the decomposition is shown
in Fig.(\ref{edge}).

\begin{figure}[h]
\begin{center}
\includegraphics[angle=0,width=8cm,height=4cm]{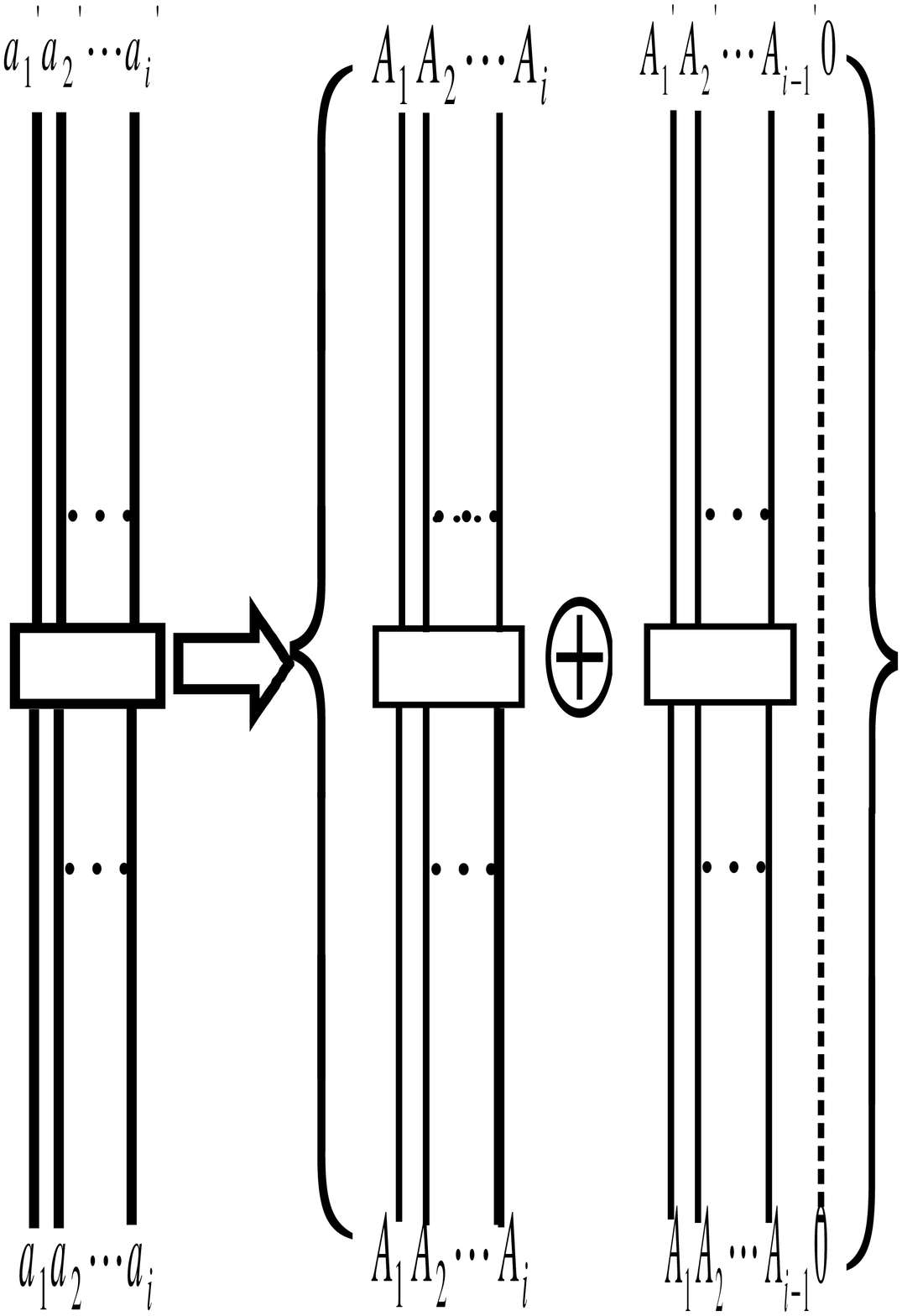}
\caption{The rope components of general edges} \label{intro11}
\end{center}
\end{figure}

Next we consider the vertex and intertwiners in spin networks. In
context of representation theory, it's equivalent to concern the
tensor product of irreducible representations. we note that the
only Lie superalgebras for which all finite dimensional
representations are completely reducible are the direct products
of $Osp(1|2n)$ superalgebras and semi-simple Lie algebras
 (Djokovi$\acute{c}$-Hochschild theorem\cite{DH}). In context of spin
networks, we will have no trouble to connect the edges together at
the same vertex. we only need to find out the appropriate
admissible conditions and label the vertex by the correct
intertwiners, which corresponds to finding the Clebsch-Gordan
series for the tensor products of the irreducible representations.
To be able to do that, the possible way is to decompose the edge
into the direct sum of ones in ordinary $Sp(2n)$ spin networks
since the tensor product of two $Sp(2n)$ representations are
discussed and the $CG$ coefficients are computable\cite{Young}.
Next let us go back to consider the decomposition of super edges
into the ordinary $Sp(2n)$ edges. In the case that only
symmetrizations involve, the edge will decomposed into two
components, which is illustrated in fig.(\ref{intro11}).

\begin{figure}[h]
\begin{center}
\includegraphics[angle=0,width=8cm,height=4cm]{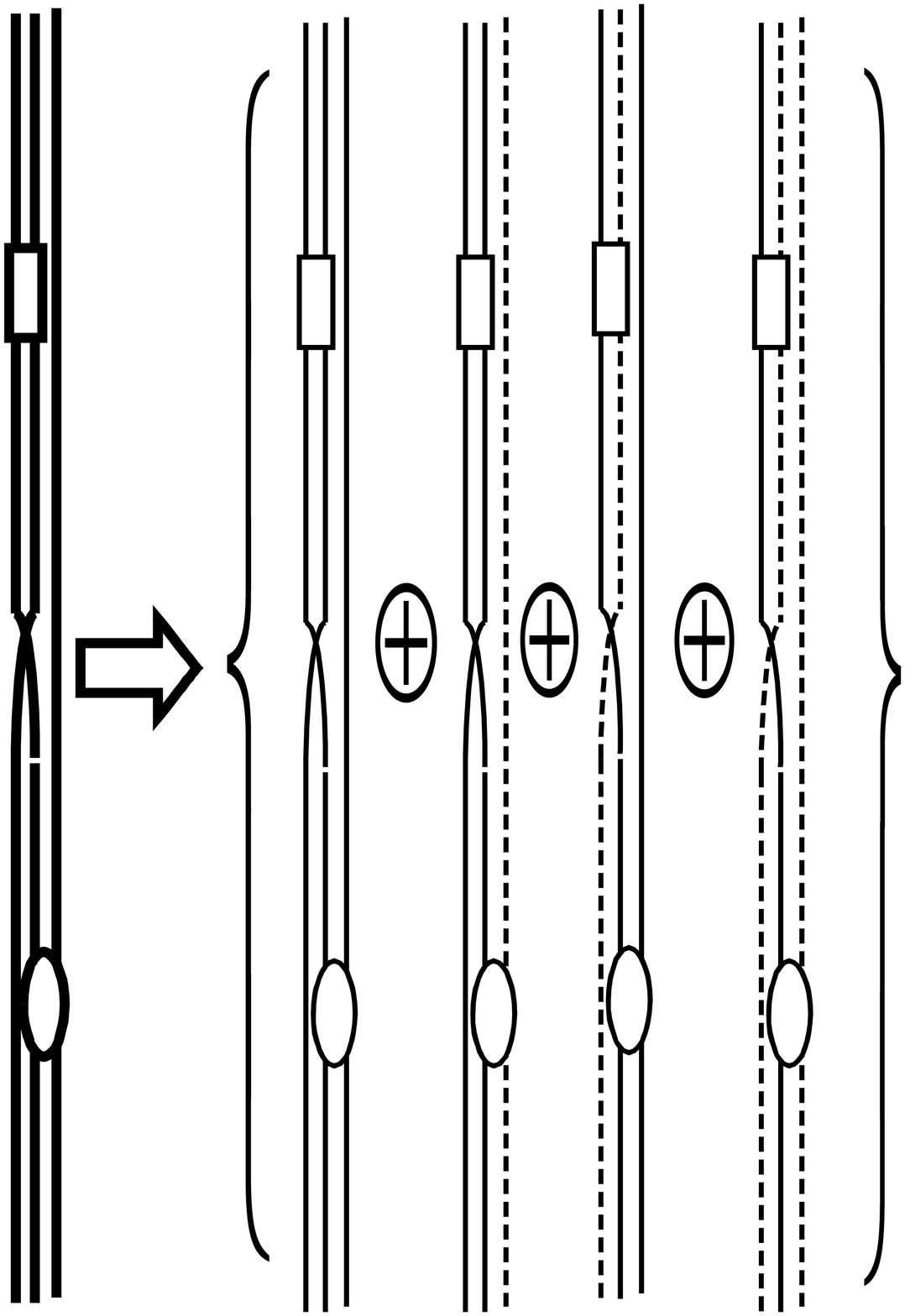}
\caption{Decomposition of super spin network into ordinary
$Sp(2n)$ ones}\label{dec}
\end{center}
\end{figure}

However, if both symmetrization and anti-symmetrizations are
involved, then we would find the terms will be more than two, and
the specific calculation is possible to carry out. Normally we
have the following strategy to obtain all the terms in ordinary
$Sp(2n)$ spin networks. The first term in $Sp(2n)$ spin networks
has the same shape as the super one, then we divide the ropes into
symmetrized groups in which each rope is symmetrized with one
another, then pick out {\em at most} one rope from each
symmetrized group in turn and replace them by the dotted lines
such that will get all the graphs in terms of the ordinary
$Sp(2n)$ spin networks. Finally we still need kick out any graph
which contains at least two dotted but {\em antisymmetrized} lines
since it vanishes. Among all the graphs remaining, we also have to
identify some equivalent graphs and maybe it's a little
complicated work. However, in context of Young supertableaux, the
procedure is simple. We find the supertableaux of $Osp(1|2n)$ can
be decompsed into the direct sum of ordinary tableaux by removing
at most one box from each row. So maybe another practical way is
to do the decomposition in context of Young tableaux and then
transform the Young diagram into edges of the spin networks. For
example, the super one $(1, 1, 0,..., 0)$ can be decomposed into
four $Sp(2n)$ terms as shown in fig.(\ref{dec}).

Then the tensor product of two edges in super spin networks will
correspondingly be decomposed to direct sum of the tensor products
of edges in $Sp(2n)$ spin networks.

\subsection{Evaluation}

In this paper we haven't carried out a specific calculation for
the evaluation of $Osp(1|2n)$ spin networks yet. However, based on
the following analysis, we would like to claim that it is
definitely possible to do so due to the features of $Osp(1|2n)$
superalgebra. Basically we have two ways to evaluate such super
spin networks. One way is to decompose the super one into the
ordinary $Sp(2n)$ spin networks, as we argued above; the other way
is given as follows. \cite{RS} shows that there is a one-to-one
correspondence between the graded representation of $Osp(1|2n)$
and the non-spinorial representations of $O(2n+1)$. First, if a
graded irreducible representation of $Osp(1|2n)$ and a
non-spinorial irreducible representation of $O(2n+1)$ have the
same highest weight, one has \f dim(\rho^{osp(1|2n)}(a_1,...,
a_n))=dim(\rho'^{O(2n+1)}(a_1,..., a_n)).\ff Moreover, the
multiplicity of any weight is the same for both representations.
As a direct application of the argument, consider the tensor
products of both irreducible representations having the same
highest weights, then we can see the Clebsch-Gordan ($CG$) series
coincide, in particular, since the tensor products are completely
reducible, their $CG$ series can be obtained by counting the
multiplicities of their weights.

So far, the one-to-one correspondence also give us an alternative
practical way to evaluate the $Osp(1|2n)$ spin networks by
studying the $O(2n+1)$ spin networks, which should be easy to
carry out at first.

\begin{figure}[h]
\begin{center}
\includegraphics[angle=0,width=8cm,height=4cm]{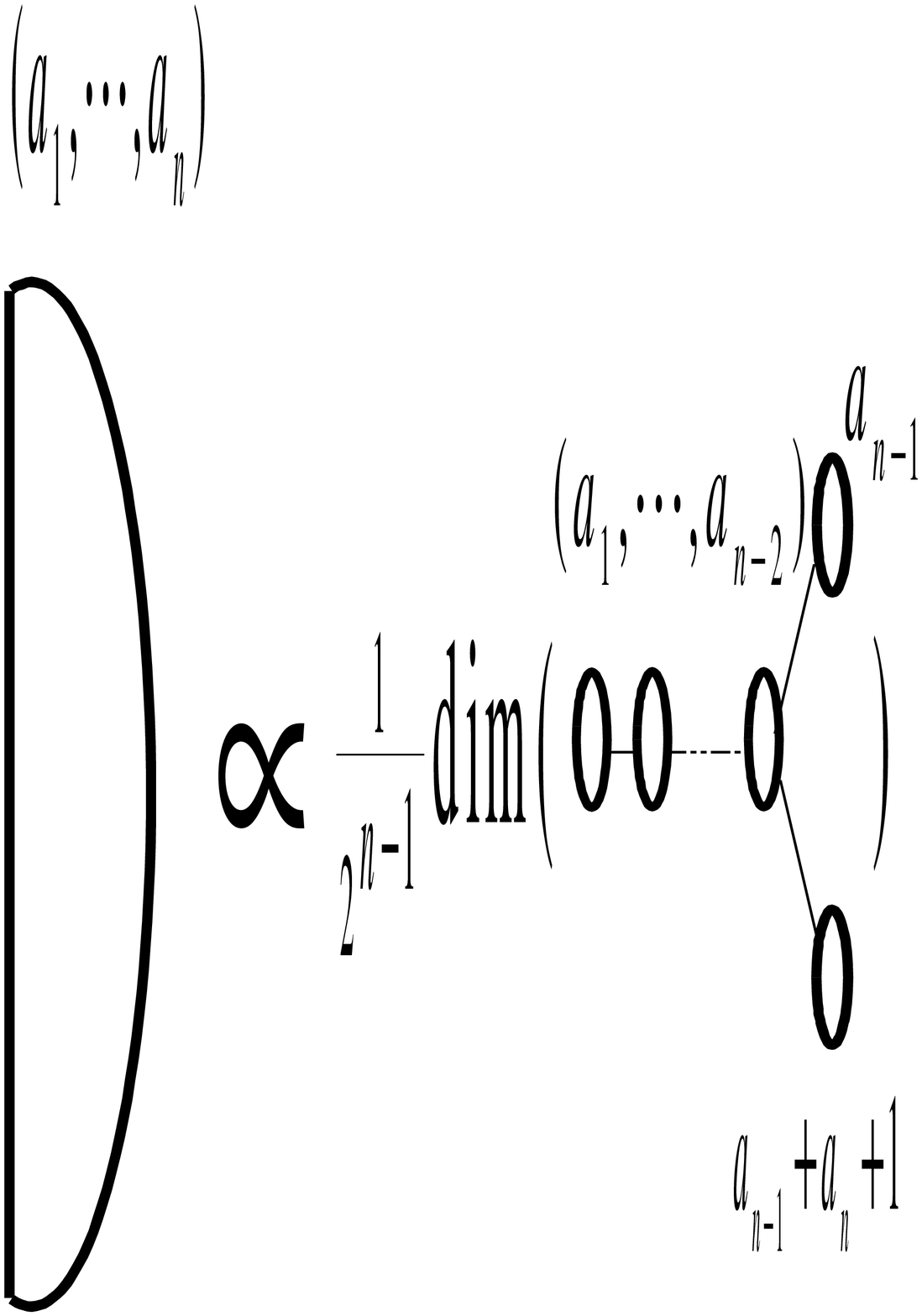}
\caption{Closure of the edges}\label{dim}
\end{center}
\end{figure}

In the last part of this section, we conjecture some examples that
could be worked out in the future. One is the closure of any edge,
namely, the supertrace of the unit element of supergroup
$Osp(1|2n)$ in such finite irreducible representation. Note that
the supertrace of supergroups is defined as\f
Str(U^{\alpha}_{\beta}):=Tr(A)-Tr(D),\ff  where we suppose the
matrix representation of the supergroup has a structure,\f
U=\left(\begin{array}{cc}
  A & B \\
  C & D
\end{array}\right).\ff
This can be done by calculating the dimensions of the Lie algebra
$O(2n+1)$ shown in fig.(\ref{dim}).

\begin{figure}[h]
\begin{center}
\includegraphics[angle=0,width=8cm,height=4cm]{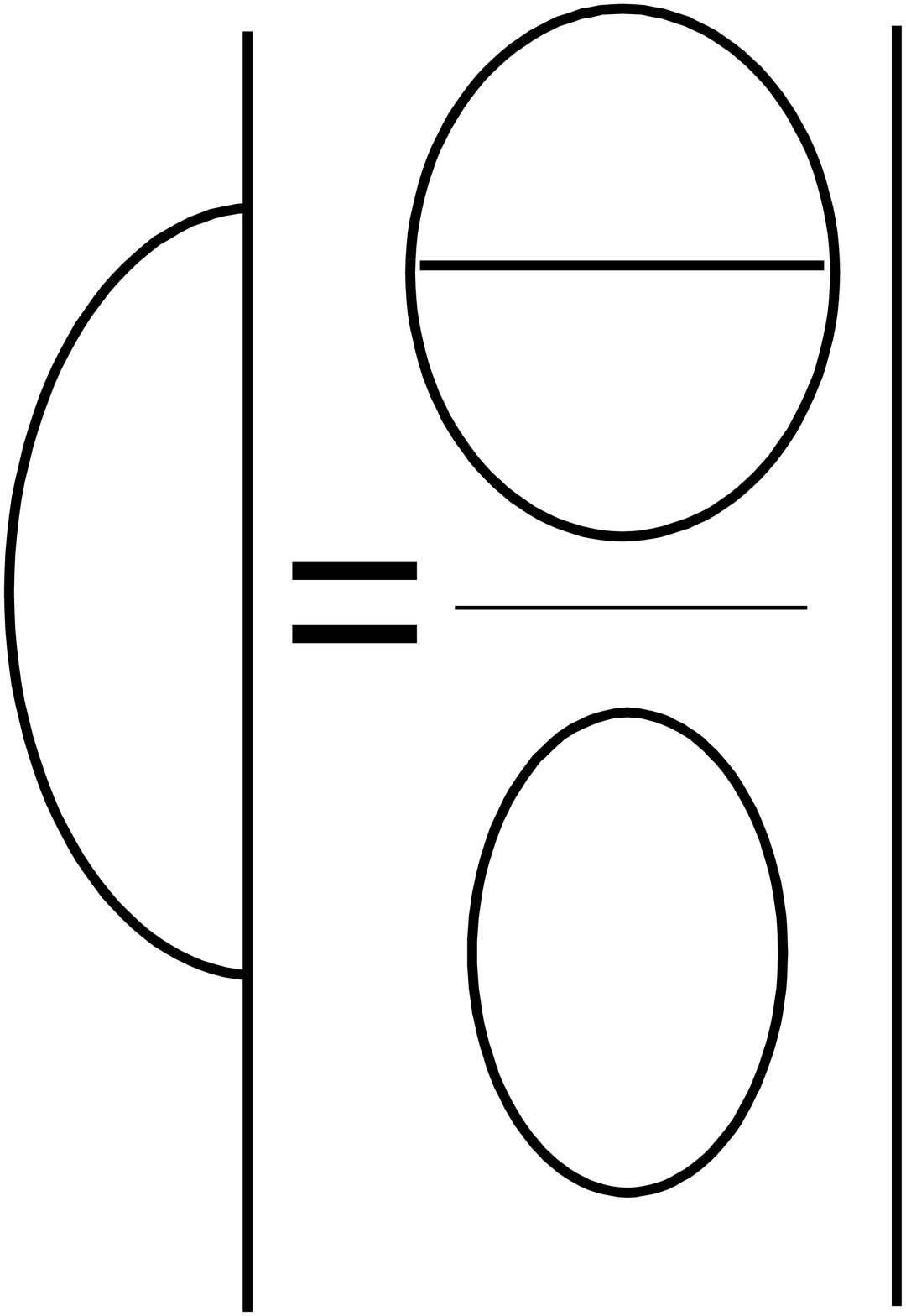}
\caption{Identity}\label{iden2}
\end{center}
\end{figure}

We find the closure of the edge is non-zero number. Thus we could
expect the identity drawn in fig.(\ref{iden2}) is evident.
They would also be applied to the non-perturbative quantization of
eleven dimensional supergravity when we consider the actions of
the operators. For instance, if we consider the quadratic Casimir
operators and their eigenvalues\footnote{There are higher-order
Casimir Operators in $Osp(1|2n)$.}. In loop quantum gravity, this
kind of operators is related to the area observable. \f
C_k(\Lambda)=\frac{1}{4n+2}(\Lambda|\Lambda+2\rho).\ff where $(|)$
is defined as the inner product of the weights and $\rho$ is
defined as, \f \rho=\omega_1+\omega_2 ...+\omega_n.\ff

The supertrace not being zero has more important meaning when we
define super spin network states and form the Hilbert space. This
good feature will allow us to define generalized
``Ashtekar-Lewandowski'' $(AL)$ measure on the space, moreover, to
show the spin networks states form a basis of the space.
\subsection{Hilbert Space}

In this part we begin with the integration theory on supergroup
manifold, and then apply it to spin network states.

Given a manifold $M$, we define a Lie superalgebra $Osp(1|2n)$
valued connection 1-form $A$. For instance, in eleven dimensional
space time, we can define the $Osp(1|32)$ connection as follows,
\f
A=:A_\mu(x)dx^{\mu}=A_\mu^a(x)P_adx^{\mu}+A_\mu^{ab}(x)J_{ab}dx^{\mu}+A_\mu^{a_1...a_5}(x)
Z_{a_1...a_5}dx^{\mu}+\Psi_\mu^{\alpha}(x)Q_{\alpha}dx^{\mu},\label{connection}\ff

where $x=(x_0, ..., x_{10})$ are local coordinates on $M$, and
$\mu$ is space time index and $a, b$ are internal indices.
$\alpha$ is spinor index. $P_a, J_{ab}, Z_{a_1...a_5}$, and
$Q_{\alpha}$ are $Osp(1|2n)$ generators while $J_{ab}$ and
$Z_{a_1...a_5}$ are skew symmetric. In (\ref{connection}) we can
think of all the components of connection $A$ as the smooth
function on the manifold $M$. Let us denote the space of smooth
connections on $M$ as $\cal A\mit$, then we define the space of
continuous functionals on $\cal A \mit$ as $Fun(\cal A\mit)$. Now
based on the linear vector space, we want to define a Hilbert
space, $\cal L\mit^2(\cal A\mit)$, namely we need introduce some
inner product of the quantum states which are the element in
$Fun(\cal A\mit)$, and then consider the completion of $Fun(\cal
A\mit)$. In loop quantum gravity, we carry out all the procedures
by introducing a special kind of functions of the connection,
which is called cylindrical functions. They are defined as the
functions of holonomies of the connection. Therefore they are
functions of group manifold as well. Then the inner products can
be defined by means of taking the integration on the group
manifold, as we know, a unique, both left- and right-invariant
measure can be defined on it.

Now we extend all the construction mentioned above to the
supersymmetric case of $Osp(1|2n)$. First note the super loop
variables can be defined by means of the holonomy of super
connections $\cal A\mit$, for details we refer to
\cite{superloop}. \f U[\cal A\mit, \gamma](s):=\cal P\mit
exp\int_{\gamma}ds \gamma^\mu A_{\mu}(\gamma(s)).\ff

Note that $U$ is an element of supergroups $Osp(1|2n)$. The loop
states can be defined as

\f \Psi_\gamma (A):=StrU[A, \gamma],\ff

where $\gamma$ is a loop in space time manifold with
$\gamma(0)=\gamma(1)$. Though the holonomy is not gauge invariant,
but their supertrace is gauge invariant indeed.

It's well known that there is a probability measure, Haar measure,
for the compact Lie groups such that we could define a unique
normalized both left- and right-invariant integration, namely,
 \f \int_G dg=1,\ \ \ \int dg f(g)=\int dg f(g_0gg_1)=\int dg
 f(g^{-1}),\ff
where $g_0, g_1$ are any group elements and $f(g)$ is also an
arbitrary function of $g$. The generalization of Haar integral for
Lie supergroups was discussed in \cite{SuperHaar}. We refer to
that paper for more details on the integration theory on
supermanifold. Here we point out that in particular case of
$Osp(1|2n)$, we can also define a generalized Haar measure on the
space of functions on supergroup which are both left- and
right-invariant\footnote{This is also true for all the semisimple
Lie supergroups\cite{SuperHaar}, however maybe it does not hold
for other supergroups. }. Therefore it's possible to develop a
positive integration theory on the space of connection, $\cal
A\mit$. Here we define the generalized Haar measure on supergroup
$Osp(1|2n)$ as: \f \int_Gdg=1, \ \ \
\int_Gg^{(\rho)}_{\alpha\beta}dg=0, \rho\neq 0.\ff

Next we consider the definition of inner product of spin network
states. Associate to every spin network, we can define a
corresponding spin network state in the Hilbert space. To show
that the set of spin network states does form a basis for the
state space in the case of $Osp(1|2n)$, we carry out the following
procedures. In spin networks we know a connection is simply to
assign a group element, $U_e$ to each edge of the graph $e$ by
taking the holonomy of the connection $A$ in the irreducible
presentation $\rho_e$ along the edge. In hence, the space of
connections in context of spin networks is\f \cal A\mit\cong
\bigotimes_{e_i} G^{e_i},\ff which is the finite product of group
$G$ and the measure is defined as\f\cal DA\mit=\bigotimes_{e_i}
dU_{e_i}.\ff We call this measure the generalized
Ashtekar-Lewandowski measure. Now we could define the inner
product of two spin network states by means of cylindrical
functions.

Consider a spin networks $\Gamma(e_i, v_j)$, we define the
cylindrical functions, $\Phi_{f_\Gamma}$, which only depend on the
holonomies of the connection $A$. Using the generalized Haar
measure on $[G]^n$. we then define the scalar product of two
cylindrical functions as,  \f\langle
\Phi_{f_\Gamma}|\Phi_{g_\Gamma}\rangle:=\int_{G^n}dU_1...dU_i\overline{f_{\Gamma}(U_1,...,
U_i)}g_{\Gamma}(U_1, ..., U_i).\ff

The Hilbert space on which $\rho_l$ is defined can be denoted as
$\cal H\mit_{\rho_l}$. Hence the total Hilbert space associated to
the spin networks can be defined as the tensor product of these
spaces, \f \cal H\mit=\bigotimes_{v_j}\cal
H\mit^{v_j}=\bigotimes_{v_j}(\bigotimes_{e_i}\cal
H\mit_{e_i})^{v_j}.\ff where $e_i$ are edges meeting at the same
vertex $v_j$.  In case of $Osp(1|2n)$, since any products of
finite dimensional irreducible representations are completely
reducible, namely, \f
\rho_1\bigotimes...\bigotimes\rho_i=\bigoplus_j v^j\rho_j\ff
correspondingly we find the tensor products of the Hilbert spaces
can be decomposed into the direct sum of Hilbert spaces on which
the irreducible representation of $Osp(1|2n)$ are defined, \f \cal
H\mit_{e_1}\bigotimes...\bigotimes\cal
H\mit_{e_i}=\bigoplus_{\rho_j}K^j\cal H\mit_{\rho_j}.\ff

Furthermore we can decompose the Hilbert space as the direct sum
of the functions on all the irreducible representations and its
conjugate: \f \cal L\mit^2(\cal A\mit / \cal G\mit)=\bigoplus_j
V_j\bigotimes V^{*}_j. \ff

Next to show that the spin network states are orthogonal and
linear independent, we exploit th generalized Peter-Weyl theorem:

Theorem:  let $\rho$ be the irreducible representation of
$Osp(1|2n)$ with the highest weights $\rho^i$, and let
$U^\rho_{\alpha\beta}$, $\alpha, \beta=1,2,...d_i(d_i=dim\rho^i)$,
be the matrix element of $\rho^i$, then:

1. \f
Fun(A)=\bigoplus_i\bigoplus^{d_i}_{\alpha,\beta=1}U^i_{\alpha\beta},\label{Peter1}\ff

2. \f\int
U^i_{\alpha\beta}\tilde{U}^j_{\gamma\sigma}(-1)^{\beta\gamma+\alpha+\beta}
=\delta_{\alpha\gamma}\delta_{ij}\frac{U^i_{\sigma\beta}}{Sdim(i)},\label{Peter2}\ff

3. \f\int
U^i_{\alpha\beta}\tilde{U}^j_{\gamma\sigma}(-1)^{\beta\gamma}
=\delta_{\beta\sigma}\delta_{ij}\frac{U^i_{\alpha\gamma}}{Sdim(i)}.\label{Peter3}\ff
>From the formula above, we note that the non-zero supertrace plays
an important role. However for the supergroups whose supertrace
vanishes, then we would maybe have some trouble to find the
generalized Peter-Weyl theorem.

Now making use of equations (\ref{Peter2}) and (\ref{Peter3}) in
the generalized Peter-Weyl theorem, and following the procedures
in \cite{sn-qg}, it's straightforward to show the $Osp(1|2n)$ spin
states are orthogonal, \f \langle \Gamma, e^i, v_j|\Gamma',
e^{i'}, v_{j'}\rangle=
\delta_{\Gamma\Gamma'}\delta_{ii'}\delta_{jj'}.\ff At the same
time, using the equation (\ref{Peter1}) we can show the set of
spin network states is completed such that any state in the
Hilbert space can be expressed into the sum of the $Osp(1|2n)$
spin network states, \f |\Phi\rangle=\sum_iC_i|\Gamma_i\rangle.\ff
Therefore the spin network states do form a basis for the Hilbert
space $\cal L\mit^2(\cal A\mit / \cal G\mit)$.

Finally we point out that in the case of quantum deformed
superalgebra, we could also construct the q-deformed spin networks
and the corresponding spin network states, which will have
important application when we study the casual evolution of the
spin networks and supersymmetric spin foams. In the q-deformed
case, the ordinary sum of two superspins $J_1+J_2$ can not give a
third superspin any more when q is not equal to one. The notion of
co-product has to be introduced. The generalized Peter-Weyl
theorem for $U_q(osp(1|2n))$ is given in $\cite{Haar}$.

In the class of $Osp(1|2n)$ spin networks, we are particularly
interested in the case of $n=16$, namely $Osp(1|32)$, since this
superalgebra is related to the eleven dimensional supergravity
\cite{fda,dAF,D=11} and M theory\cite{M-theory}. As we know, in
eleven dimensional supergravity, the Superpoincare algebra with
two and five form central charges are obtained by taking a
Inonu-Wigner contraction of $Osp(1|32)$. This superalgebra has
many facets in different dimensions which is studied in
\cite{osp32}. As a result, we expect the $Osp(1|32)$ spin networks
will take its own advantages when we try to carry out a background
independent and non-perturbative quantization of the eleven
dimensional supergravity, and M theory.

\section{$Osp(N|2)$ spin networks}

In previous section we have described $Osp(1|2n)$ spin networks
following the proposal we give in overview. In this section we
will discuss $Osp(N|2)$ spin networks briefly. We are interested
in this kind of spin networks because they are related to the
chiral supergravities\cite{super}. But in this class the case of
$N=2$ is different from the others since the even part of this
superalgebra is $SO(2)\times Sp(2)$ such that the first number
$a_1$ of Dynkin labels can be any complex number, however for the
others it has to be non-negative integer. Its construction and
application to supergravity are studies in
\cite{supern=2,holon=2}. Here we consider the general case when
$N$ is larger than two.

The basis vectors of the fundamental representation of $Osp(N|2)$
span a $N+2$ dimensional vector space. Similar to the case of
$Osp(1|2n)$, its higher finite dimensional irreducible
representations can be obtained by standard symmetrization and
anti-symmetrization procedures on fundamental representation.
Therefore, in context of spin networks, we can decompose the edge
into ropes.

\begin{figure}[h]
\begin{center}
\includegraphics[angle=0,width=8cm,height=4cm]{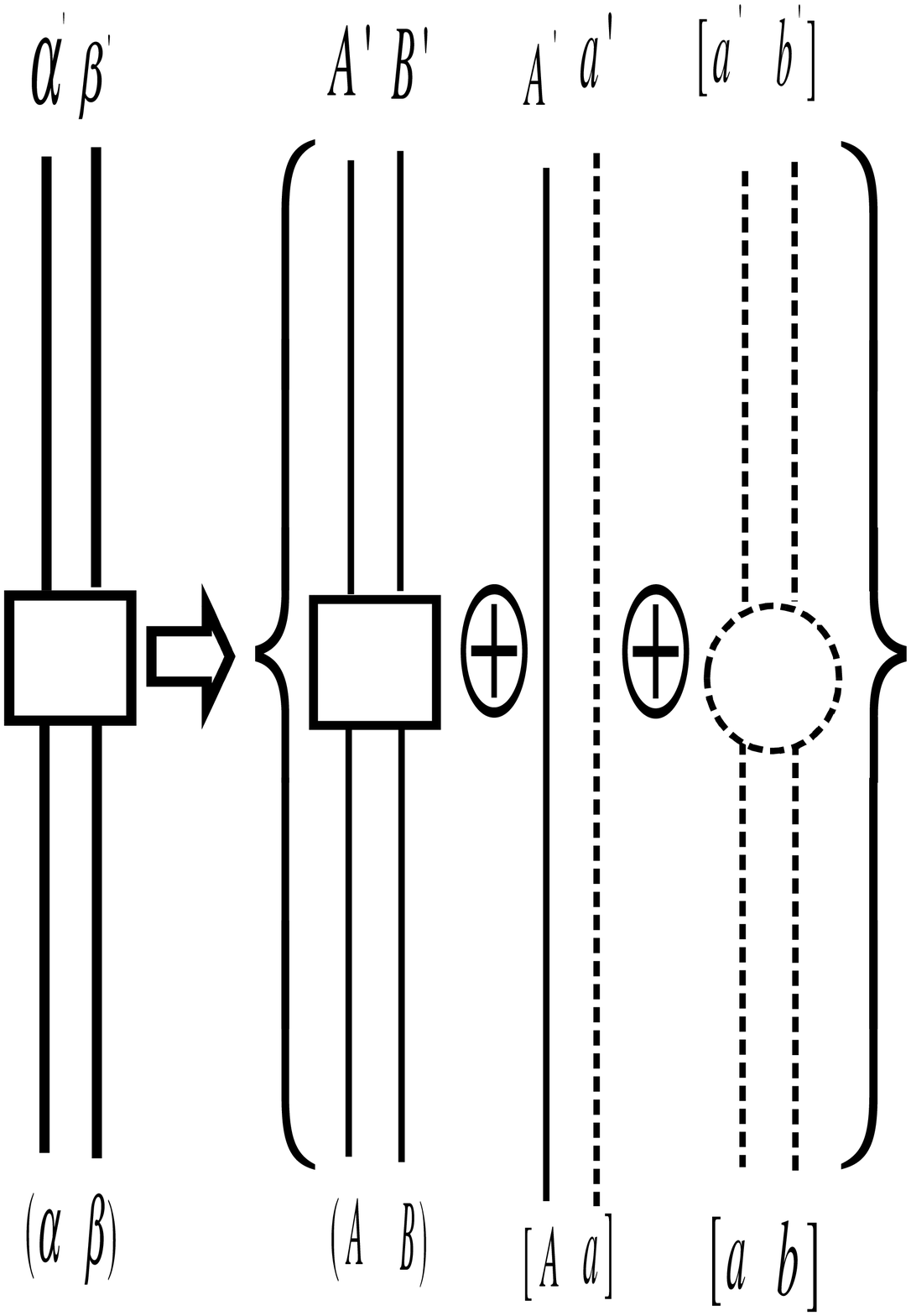}
\caption{Symmetrizer with color two}\label{intro9}
\end{center}
\end{figure}

\begin{figure}[h]
\begin{center}
\includegraphics[angle=0,width=8cm,height=4cm]{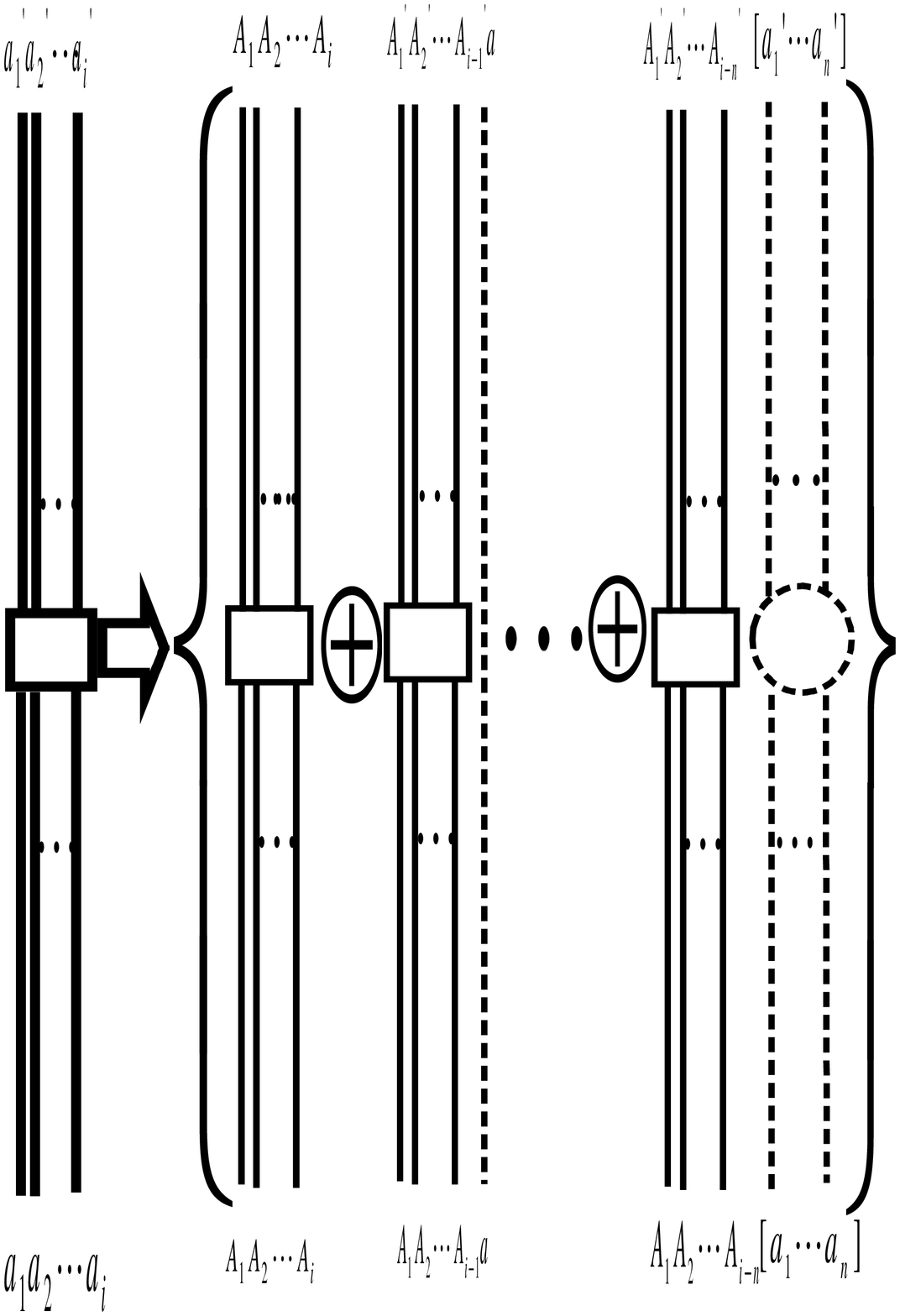}
\caption{Symmetrizer with color $(i, 0, ..., 0)$}\label{intro10}
\end{center}
\end{figure}

In fact any representation of basic Lie superalgebra can be
decomposed into the direct sum of irreducible representations of
the even subalgebra, which means any edge in super spin networks
can be decomposed into sum of normal edges which labeled by the
irreducible representations of the even subalgebra. In case of
$Osp(N|2)$, we can decompose the $Osp(N|2)$ spin networks into the
direct sum of the ordinary spin networks with $SO(N)\bigotimes
Sp(2)$. In particular, if only the symmetrization procedures are
considered, we have simple decompositions. Fig.(\ref{intro9})
illustrates the basic one which has the color $(2,0,...,0)$.
Since the $SO(N)$ indices are antisymmetrized, the decomposition
of the graphs terminate at the term which contains $N$ dotted
lines, see fig.(\ref{intro10}).

However, unlike the case of $Osp(1|2n)$, some questions arise in
$Osp(N|2)$ case as we try to construct the Hilbert space based on
spin network states.  As we mentioned before, both typical and
atypical representations are present for other superalgebras
except $Osp(1|2n)$. The key difference of these two types of
representations is that all the typical representation is either
irreducible or completely reducible, however, the atypical one
maybe is not completely reducible (i.e., reducible but not
decomposable). In group theory, there is a simple meaning of
complete reducibility, namely the representation space should be a
Hilbert space and the representation should be unitary.
Correspondingly in context of spin networks we will meet some
serious problems to show the corresponding spin net work states
form a basis for the Hilbert space. In this case, we don't know
there exists some kinds of generalized Peter-Weyl theorem, though
the left-invariant Haar integral for such supergroups are also
discussed in \cite{SuperHaar}. But Let us note there are some
classes, which are called star representations and graded star
representations in irreducible representations of the
superalgebras\cite{dic, Osp(2/2)b}. Inside each class the complete
reducibility is reserved indeed and the CG series can be given
also. Thus, spanning spin network states in such subspaces, we
still expect they could be well defined to form a basis for such
Hilbert spaces. Note that in quantum gravity, only the
$\it{physical}$ Hilbert space is what we want finally. So it's
quite interesting to find some self-consistent subset of spin
networks which are well defined and investigate the possible
applications to quantum theory.

In the end of this section, let us make a summation by listing the
correspondence between the construction of spin networks and the
representation theory of the superalgebra.\\

\begin{tabular}{|c|c|}\hline
  Spin networks & Representation theory of superalgebra \\
  Edges &  Irreducible representation\\
  Vertex & Intertwiner operator\\
  Ropes & Fundamental representations \\
  Admissible conditions & Clebsch-Gordan series \\
  Closure of edges & Supertrace of the unit element\\
  Spin network states & Cylindrical functions\\
  Inner product of spin network states & Generalized Haar
  measure\\
  Orthogonality and linear independence of states & Generalized
  Peter-Weyl theorem\\ \hline
\end{tabular}

\section{Discussions}

In this paper we have presented a general introduction to the
construction of supersymmetric spin networks. All the strategy in
ordinary $SU(2)$ spin networks can be employed to the case of
$Osp(1|2n)$ spin networks. In particular the spin network states
form a basis for the Hilbert space $\cal L\mit^2(\cal A\mit/\cal
G\mit)$. But normally it's becoming more complicated to carry out
a practical way to evaluate the spin network graphs, as in general
both symmetrization and antisymmetrization procedures are involved
in higher finite dimensional irreducible representations. However,
it's definitely plausible to give a specific calculation on
evaluation of graphs in the future by following strategies given
in present paper. On the other hand, due to the properties of the
superalgebra, it's unclear if we could construct spin network
states as the basis of Hilbert space based on any kind of
superalgebra. Here we only discussed the $Osp(N|2)$ spin networks
briefly and found this problem should be concerned seriously.

To apply spin-network techniques to quantum supergravities and
Yang-Mills theory is very promising and important. At root they
are powerful tools to quantize gauge theories along the
non-perturbative and background independent approach.

First we expect the corresponding spin network states would be
applied to construct the Hilbert space of quantum supergravities.
In this framework we could find a practical way to consider the
action of the operators, and calculate the spectra of the
observables. In \cite{superspin}, we have made a first step to
calculate the spectrum of area operator in $N=1$ Chiral
supergravity, and the extension to the case of $N=2$ is also done
in \cite{supern=2}.

Second we propose the holographic hypothesis \cite{linking,
hologr} can be testified in the framework of non-perturbative
quantum supergravities. This conjecture has been testifies in N=1
and N=2 supergravity respectively \cite{holon=2,holon=1}. It is
the supersymmetric spin networks that make it possible to count
the number of degrees of freedom on the boundary such that we find
the relations between the area of the boundary and the number of
the states do satisfy Bekenstein's conditions.

Until now we only take into account for the finite dimensional
representation of the superalgebra. It is worth studying what role
the spin networks will play if we consider the infinite
dimensional representation of the superalgebra.

\section*{Acknowledgement}

I am very grateful to my advisor, Prof. Lee Smolin, for his
contribution and encouragement during the course of this work.
Also I'd like to thank Laurent Freidel, Marcus Gaul, Yongge Ma and
Daniele Oriti for their continuous interests in this subject and
communication. I am also grateful to the theoretical physics group
at Imperial College for hospitality when part of work was done
there. This work was supported by the NSF through grant
PHY95-14240 and a gift from the Jesse Phillips Foundation.


\begin{thebibliography}{99}
\bibitem{penrose}
R. Penrose, in {\em Quantum theory and beyond}, ed T Bastin,
Cambridge U Press, 1971.
\bibitem{sn-gauge}
J. Kogut and L. Susskind, Phys. Rev. D11 (1975)395; W. Furmanski
and A. Kowala, Nucl. Phys. B291(1987)594. J. Baez, Adv.Math. 117
(1996) 253.
\bibitem{sn-topo}
E. Witten, Commum. Math. Phys. 121(1989)351; V. Turaev and O.
Viro, Topology 31(1992) 865; H. Ooguri, Mod. Phys. Lett
A7(1992)2799; L. Crane and D. Yetter, in {\em Quantum Topology},
eds L. H. Kauffman and R. A. Baadhio, World Scientific Press,
(1993) 120; L. Crane and I. B. Frenkel, J. Math. Phys.
35(1994)5136; L. Crane and D. Yetter, {\em On algebra structure
implicit in topological quantum field theories}, Kansas preprint
1994; T. J. Foxon, Class.Quant.Grav. 12 (1995) 951; C. Rovelli,
Phys. Rev. D48 (1993) 2702; S. Mizoguchi and T. Tada, Phys. Rev.
Lett. 68 (1992) 1795.

\bibitem{sn-qg}
C. Rovelli and L. Smolin, Phys. Rev. D52 (1995) 5743; J. Baez,
{\em Spin networks in nonperturbative quantum gravity}, in The
Interface of Knots and Physics, ed. Louis Kauffman, A.M.S.,
Providence, 1996, 167;  A. Ashtekar, J. Lewandowski, D. Marolf, J.
Mourao, T. Thiemann, J. Math. Phys. 36 (1995) 6456.

\bibitem{rigorous}
A Ashtekar, J Lewandowski, D Marlof, J Mour\~{a}u and T Thiemann,
{\em Quantization of diffeomorphism invariant theories of
connections with local degrees of freedom}, gr-qc/9504018, JMP 36
(1995) 519; A. Ashtekar and J. Lewandowski, {\em Quantum Geometry
I: area operator}, gr-qc/9602046; J. Lewandowski, {\em Volume and
quantization}, gr-qc/9602035;  T. Thiemann, Quantum Spin Dynamics
I-VI, Class. Quant. Grav. 15 (1998) 839-873, 875-905, 1207-1247,
1249-1280, 1281-1314, 1463-1485, 1487-1512. gr-qc/9606092,
gr-qc/9606089, gr-qc/9606090, gr-qc/9705020, gr-qc/9705021,
gr-qc/9705019, gr-qc/9705018, gr-qc/9705017.

\bibitem{sn1}C. Rovelli and L. Smolin
{\it Discreteness of area and volume in quantum gravity}
 Nuclear Physics B 442 (1995) 593.  Erratum: Nucl. Phys.
B 456 (1995) 734.

\bibitem{fotini}
F. Markopoulou, {\it Dual formulation of spin network evolution}
preprint, March 1997, gr-qc/9704013; F. Markopoulou and L. Smolin
{\it Causal evolution of spin networks}  gr-qc/9702025. CGPG
preprint (1997), Nuclear Physics B, Nucl.Phys. B508 (1997)
409-430; F. Markopoulou and L. Smolin {\it Quantum geometry with
intrinsic local causality}, Phys. Rev. D58 (1998) 084032,
gr-qc/9712067.

\bibitem{foam}M. Reisenberger,
{\it A lattice worldsheet sum for 4-d Euclidean general
relativity}, gr-qc/9711052; M. Reisenberger and C. Rovelli, {\it
``Sum over Surfaces'' form of Loop Quantum Gravity}, Phys.Rev. D56
(1997) 3490, gr-qc/9612035; J. Baez,{\it Spin foam models},
Class.Quant.Grav. 15 (1998) 1827, gr-qc/9709052; J. Baez, {\em An
Introduction to Spin Foam Models of Quantum Gravity and BF Theory
}, gr-qc/9905087.

\bibitem{othersn}J.
Barrett and L. Crane, {\em Relativistic spin networks and quantum
gravity}, J. Math. Phys. 39 (1998) 3296, gr-qc/9709028; {\em A
Lorentzian Signature Model for Quantum General Relativity} Class.
Quant. Grav. 17 (2000) 3101, gr-qc/9904025;  S. Davids, {\em
Semiclassical Limits of Extended Racah Coefficients}, J. Math.
Phys. 41 (2000) 924, gr-qc/9807061; M. Bojowald, {\em Abelian
BF-Theory and Spherically Symmetric Electromagnetism }, J. Math.
Phys. 41 (2000) 4313, hep-th/9908170.

\bibitem{superspin}
Y. Ling and L. Smolin, {\em The supersymmetric spin networks and
quantum supergravity}, Phys. Rev. D61, 044008(2000),
hep-th/9904016.

\bibitem{supern=2}
Y. Ling and L. Smolin, {\em $N=2$ quantum supergravity}, preprint
in preparation.

\bibitem{KL}
L. Kauffman, S. Lins, {\em Temperley-Lieb Recoupling Theory and
Invariants of 3-Manifolds}, Princeton U Press, 1994.

\bibitem{Kac}
V. Kac, Adv. Math.26, (1977)8; Commun. Math. Phys. 53(1977)33;
Lecture Notes in Mathematics vol.676( New York, Springer).

\bibitem{dic}
L. Frappat, A. Sciarrino and P. Sorba, {\em Dictionary on Lie
Superalgebras}, hep-th/9607161.

\bibitem{RS}
V. Rittenberg and M. Scheunert, {\em A Remarkable Connection
Between the Representations of the Lie Superalgebras osp(1, 2n)
and the Lie Algebra O(2n+1)}, Commun. Math. Phys. 83, 1 9(1982).

\bibitem{GSU5}
A. Balantekin and I. Bars, {\em Dimension and character formulas
for Lie supergroups}, J. Math. Phys.22,(1981)1149; {\em
Representation of supergroups}, J. Math. Phys.22,(1981)1810.

\bibitem{DH}
G. Hochschild, III, J. Math. 20, 107(1976); D.Djokovi$\acute{c}$
and G. Hochschild, III, J. Math.20, 134(1976); D.
Djokovi$\acute{c}$, J. Pure Appl. Alg. 7, 217(1976).

\bibitem{Young}
B. More, A. Sciarrino and P. Sorba, {\em Representation of the
$Osp(M|2n)$ and Young supertableaux}, J. Phys. A: Math. Gen.
18(1985)1597. G. Girardi, A. Sciarrino and P. Sorba, {\em
Kronecker product of Sp(2n) representations using generalised
Young tableaux}, J. Phys. A: Math. Gen. 16(1983)2609.

\bibitem{superloop}L. Urrutia, {\em Towards a loop representation of
connection theories defined over a super Lie algebra}, aipproc
style Based on the lectures given at the Fifth Workshop on
Particles and Fields, Puebla, Nov. 1995, hep-th/9609001.

\bibitem{SuperHaar}D. Williams and J. F. Cornwell, {\em Haar
Integral for Lie supergroups}, J. Math. Phys. 25(1984)2922.

\bibitem{Haar} H. C. Lee and R. B. Zhang, {\em Geometry and
representations of the quantum supergroup $Osp_q(1|2n)$},
math.QA/9804111; M. Scheunert, R. B. Zhang {\em Invariant
integration on classical and quantum Lie supergroups},
math.RA/9911200.

\bibitem{fda}L.~Castellani, P.~Fre and P.~van Nieuwenhuizen,
``A Review Of The Group Manifold Approach And Its Application To
Conformal Supergravity,'' Annals Phys.\  {\bf 136}, 398 (1981).

\bibitem{dAF}R.~D'Auria and P.~Fre,
``Geometric Supergravity In D = 11 And Its Hidden Supergroup,''
Nucl.\ Phys.\  {\bf B201}, 101 (1982). Erratum-ibid.B206:496,1982.

\bibitem{D=11}
Y. Ling and L. Smolin, {\em Eleven dimensional supergravity as a
constrained topological field theory}, hep-th/0003285.

\bibitem{M-theory}
P. K. Townsend, {\em M theory from its superalgebras}, Cargese
lectures 1997, hep-th/9712004; M. Gunaydin, {\em Unitary
supermultiplets of $Osp(1/32,R)$ and M theory}, Nucl.Phys. B528
(1998) 432-450, hep-th/9803138.

\bibitem{osp32}E. Bergshoeff,
A. Proeyen, {\em The many faces of $OSp(1|32)$},
 Class.Quant.Grav. 17 (2000)3277, hep-th/0003261.

\bibitem{super}T. Jacobson, {\it New variables for canonical
supergravity} Class. Quant. Grav. 5 (1988) 923; D. Armand-Ugon, R.
Gambini, O. Obregon, J. Pullin, {\it  Towards a loop
representation for quantum canonical supergravity},
hep-th/9508036, Nucl. Phys. B460 (1996) 615; H. Kunitomo and T.
Sano {\it The Ashtekar formulation for canonical N=2
supergravity}, Prog. Theor. Phys. suppl. (1993) 31; Takashi Sano
and J. Shiraishi, {\it The Non-perturbative Canonical Quantization
of the N=1 Supergravity}, Nucl. Phys. B410 (1993) 423,
hep-th/9211104; {\it The Ashtekar Formalism and WKB Wave Functions
of N=1,2 Supergravities}, hep-th/9211103; K. Ezawa, {\it
Ashtekar's formulation for N=1, N=2 supergravities as constrained
BF theories},  Prog. Theor. Phys.95:863-882, 1996,
hep-th/9511047.T. Kadoyoshi and S. Nojiri, {\it N=3 and N=4 two
form supergravities}, Mod. Phys. Lett. A12:1165-1174,1997,
hep-th/9703149; L. F. Urrutia {\it Towards a loop representation
of connection theories defined over a super-lie algebra}
hep-th/9609010.

\bibitem{holon=2}
Y. Ling and L. Smolin, {\em Holography, BPS states and N=2 quantum
supergravity}, preprint in preparation.

\bibitem{Osp(2/2)b}
M.Scheunert, W.Nahm, and V.Rittenberg, {\em Graded Lie Algebras:
Generalization of Hermitian representations}, J. Math.
Phys.18(1977)146; {\em Irreducible representations of the osp(2,1)
and spl(2,1) graded Lie algebras}, J.Math.Phys.18,(1977)155.

\bibitem{linking}
L. Smolin, {\em Linking topological quantum field theory and
nonperturbative quantum gravity},gr-qc/9505028, J. Math. Phys.
36(1995)6417;

\bibitem{hologr}
L. Smolin, {\em A holograpic formulation of quantum general
relativity}, hep-th/9808191.

\bibitem{holon=1}
Y. Ling and L. Smolin, {\em Holographic formulation of $N=1$
supergravity}, preprint in preparation.

\end{thebibliography}
\end{document}